\documentclass[useAMS,usenatbib]{mn2e}
\usepackage{mathptmx}
\usepackage{txfonts}
\usepackage[T1]{fontenc}

\usepackage{graphicx}

\newcommand\msol{{\cal M_{\odot}}}
\newcommand\teff{{T_{\rm eff}}}
\newcommand\vsini{{v\,\sin\,i}}
\newcommand\lta{\mathrel{\hbox{\raise 0.6 ex \hbox{$<$}\kern
                   -1.8 ex\lower .5 ex\hbox{$\sim$}}}}
\newcommand\gta{\mathrel{\hbox{\raise 0.6 ex \hbox{$>$}\kern
                   -1.7 ex\lower .5 ex\hbox{$\sim$}}}}

\title[ZAHB Stellar Models]{Zero-Age Horizontal Branch Models for $-2.5
\le$ [Fe/H] $\le -0.5$ and their Implications for the Apparent Distance Moduli
of Globular Clusters}

\author[D. A. VandenBerg]{Don A.~VandenBerg \thanks{E-mail: vandenbe@uvic.ca } 
\\
Department of Physics and Astronomy, University of Victoria, 
P.O. Box 1700, STN CSC, Victoria, BC, Canada V8W 2Y2 }

\date{Accepted XXX. Received YYY; in original form ZZZ}
\pubyear{2022}

\begin{document}
\label{firstpage}
\pagerange{\pageref{firstpage}--\pageref{lastpage}}
\maketitle

\begin{abstract}

Grids of zero-age horizontal branch (ZAHB) models are presented, along with a
suitable interpolation code, for $-2.5 \le$ [Fe/H] $\le -0.5$, in steps of
$0.2$ dex, assuming $Y = 0.25$ and 0.29, [O/Fe] $= +0.4$ and $+0.6$, and
[$m$/Fe] $= 0.4$ for all of the other $\alpha$ elements.  The HB populations of
37 globular clusters (GCs) are fitted to these ZAHBs to derive their apparent
distance moduli, $(m-M)_V$.  With few exceptions, the best estimates of their
reddenings from dust maps are adopted.  The distance moduli are constrained
using the prediction that $(M_{F606W}-M_{F814W})_0$ colours of metal-poor, 
main-sequence stars at $M_{F606W} \gta 5.0$ have very little sensitivity to
[Fe/H].  Intrinsic $(M_{F336W}-M_{F606W})_0$ colours of blue HB stars, which
provide valuable connections between GCs with exclusively blue HBs and other
clusters of similar metallicity that also have red HB components, limit the
uncertainties of {\it relative} $(m-M)_V$ values to within $\pm 0.03$--0.04 mag.
The ZAHB-based distances agree quite well with the distances derived by
Baumgardt \& Vasiliev (2021, MNRAS, 505, 5957).  Their implications for GC ages
are briefly discussed.  Stellar rotation and mass loss appear to be more
important than helium abundance variations in explaining the colour-magnitude
diagrams of second-parameter GCs (those with anomalously very blue HBs for their
metallicities).

\end{abstract}


\section{Introduction}
\label{sec:intro}

A zero-age horizontal branch (ZAHB) typically consists of a few dozen stellar
models that predict the luminosities and effective temperatures ($\teff$s) of
stars that are just beginning their core He-burning phases of evolution.  Such
models have the same helium core mass and chemical abundance profiles above the
core, as derived from an appropriate red-giant precursor (see, e.g.,
\citealt{ptc04}, \citealt{sw05}, \citealt{vdc16}), but a range of envelope, and 
hence total, masses to take into account star-to-star variations in mass loss
along the upper red-giant branch (RGB) during the preceding evolution.
Fortunately, ZAHBs that are applicable to old stellar populations are nearly
age-independent, as shown most recently by \citet[hereafter DVKF17]{dvk17} and
\citet{phc21}.  Consequently, it is necessary to compute only a single 
evolutionary track, for a mass that has a predicted lifetime of 12 Gyr or so,
to the tip of the RGB in order to obtain the information that is needed to
generate a ZAHB for the assumed initial abundances of helium and the metals.  In
effect, ZAHBs are decoupled from isochrones; i.e., the same ZAHB can be fitted
to the HB population of a given globular cluster (GC) irrespective of the age
of the isochrone that provides the best fit to the turnoff (TO) observations.  

Because ZAHB sequences are nearly horizontal at colours that are typical of
upper main-sequence (MS) and TO stars in optical colour-magnitude diagrams 
(CMDs) of GCs, they have often been used to derive the apparent distance moduli
of these systems, particularly when they contain red HB populations (e.g.,
\citealt{van00}; \citealt{rpd05}; \citealt[hereafter VBLC13]{vbl13}).  The
fitting of observed HBs to ZAHB models has the advantage of being insensitive
to $\teff$ and reddening uncertainties, in addition to having a relatively weak
dependence on metallicity.  Most stellar evolutionary computations over the
years have found that the slope of the $M_V$(HB) versus [Fe/H] relation in the
vicinity of the instability strip is $\approx 0.18$--0.20 (\citealt{ldz90};
\citealt[and references therein]{ccd99}; \citealt{vsf00}), which means that a
0.1 dex error in the adopted metallicity results in an error of $\lta 0.02$ mag
in the derived value of $(m-M)_V$.  Provided that the ZAHB models are
trustworthy, ages that are based on the difference in magnitude between the
ZAHB and the TO, when evaluated using the methods described by VBLC13, should be
especially reliable.  (This also requires, of course, that the stellar models
are generated for best estimates of the chemical abundances.)

The main difficulty with this approach is observational, insofar as the flat
portions of ZAHBs are not populated in a large number of Galactic GCs, many of
which have metallicities in the range $-2.0 \lta$ [Fe/H] $\lta -1.5$.  When the
core He-burning stars are distributed along steep, blue tails (in optical CMDs),
small errors in the predicted colours of the stellar models or in the adopted
reddenings can easily alter the distance moduli that are obtained from fits of
the HB populations to ZAHB models by as much as 0.1--0.2 mag.  However, as shown
below, it is possible to do much better than that if other considerations are
taken into account --- notably, the near independence of the location of the
MS fiducials of the most metal-poor GCs on the
$[(M_{F606W}-M_{F814W})_0,\,M_{F606W}]$-diagram and similar CMDs.  In fact,
it turns out that ultraviolet photometry from the {\it Hubble Space Telescope
(HST)} UV Legacy Survey (\citealt{pmb15}, \citealt[hereafter NLP18]{nlp18}) can
also be used to place tight constraints on the relative reddenings and distance
moduli of GCs that have similar metallicities but very different HB morphologies
(such as M$\,$3 and M$\,$13).

Although the discovery of He abundance variations within GCs (\citealt{pvb05},
\citealt{mmp12}) has made the modeling of the entire distributions of their
HB stars on various CMDs quite challenging (see, e.g., DVKF17), the presence of
multiple stellar populations should not affect ZAHB-based determinations of
$(m-M)_V$.  Based on their survey of 57 GCs, \citet{mmr18} concluded that all of
the clusters in their sample contain (at least) two chemically distinct stellar
populations, which they refer to as 1G and 2G (i.e., first and second
generation), and that all 2G stars are helium enhanced with respect to those
classified as 1G, with enhancements that apparently vary from $\Delta\,Y \lta
0.01$ to $\gta 0.10$, where $Y$ is the mass-fraction abundance of helium.  As a
consequence of having higher $Y$, 2G HB stars near their ``zero-age" CMD
locations can be expected to be brighter, at a given colour, than their 1G
counterparts.  Thus, the faintest of the reddest HB stars should, presumably,
be well represented by ZAHB models for initial helium abundances that are close
to the primordial value ($Y_{\rm P} \approx 0.247$; see, e.g., \citealt{pcu18},
\citealt{pco20}).\footnote{Due to the effects of atomic diffusion, the
envelope He abundance after the first dredge-up on the RGB can be {\it less}
than the initial value --- by, for instance, $\Delta\,Y \approx 0.003$ in the
case of a $0.8 \msol$ stellar model for [Fe/H] $= -1.55$, as reported by
\citet{vdc16}.}

This expectation is supported by the direct spectroscopic determination of the
He abundance in NGC$\,$6752 by \citet{vpg09}, who obtained a mean value of $Y =
0.245 \pm 0.012$ for 7 HB stars that hot enough ($\teff > 8500$ K) to have
measureable photospheric helium lines, but not so hot ($\teff < 11,500$ K) that
the surface chemistry is modified by gravitational settling.   Even though
\citet{vgp12} obtained a significantly larger value of $Y$ ($0.29 \pm 0.01$)
from a similar study of the blue HB stars in M$\,$4, they argued that these
stars are likely to be helium-enhanced, and therefore members of the 2G
population of M$\,$4, because they are brighter than the faintest red HB stars
in this cluster.  In fact, the downward tilt of the M$\,$4 HB, in the direction from
blue to red, resembles that seen in NGC$\,$6362 (\citealt{bcr99}), which was
the subject of a recent study by \citet{vd18}.  The latter concluded from
their fits of ZAHB models to the non-variable HB stars on the $(B-V)_0,\,M_V$
diagram, from an analysis of the observed periods of member RR Lyrae variables,
and from full simulations of the cluster HB population, that its blue HB stars
have higher He abundances by as much as $\Delta\,Y \approx 0.03$ than those on
the red side of the instability strip.  Hence, the available evidence indicates
that ZAHBs for $Y \approx Y_{\rm P}$ should be relevant to at least some
fraction of the reddest HB stars in {\it all} GCs, even those that have
exclusively blue HB populations, such as NGC$\,$6752.

This investigation has been carried out to (i) provide new ZAHB models to
complement several of the grids of tracks for earlier evolutionary phases that
were recently made available by \citet{vec22} and \citet{van23}, and (ii) 
examine the implications of such models for the apparent distance moduli of GCs.
The computation of the ZAHBs and the methods used to determine the $(m-M)_V$
values of interest are described in \S~\ref{sec:models} and \S~\ref{sec:fits},
respectively.  The distance moduli of 37 GCs with metallicities in the range
$-2.4 \lta$ [Fe/H] $\lta -0.6$, supported by intercomparisons of the cluster HB
populations on UV-optical CMDs, are derived in \S~\ref{sec:dis}.  The possible
role of stellar rotation and mass loss in explaining the CMDs of
second-parameter clusters is discussed in \S~\ref{sec:second}.  A summary of
the main results of this study is provided in \S~\ref{sec:sum}, which also
compares the derived distance moduli with those obtained by VBLC13 and by
\citet{bv21}, who concluded from a thorough analysis of literature values,
supplemented by their own findings that utilized {\it Gaia} and various {\it
HST} observations, that their distance determinations should be accurate to
within a few percent for most clusters.

\section{The Computation of the ZAHB Sequences}
\label{sec:models}

The ZAHB models presented in this investigation were computed using the stellar
structure and evolution program described by \citet[and references
therein]{van23}, who provides complementary evolutionary tracks for the MS and
RGB phases, along with the means to produce isochrones from them.  A brief
summary of the physics that has been incorporated in the Victoria code since
2012, when it was last revised (see \citealt{vbd12}), is provided in
Table~\ref{tab:t1}.  Of the physics components listed therein,
diffusion likely has the greatest impact on ZAHB models because, in comparison
with the predictions of non-diffusive computations, it results in lower surface
He abundances in red giants after the first dredge-up by $\Delta\,Y ~\sim 0.015$,
with some dependence on mass and metallicity, leading to fainter ZAHBs by $\sim
0.05$ mag (see, e.g., \citealt{pv91}).

The Victoria code does not follow the
settling of the metals (for some discussion of this point, see \citealt[their
\S~2]{vbf14}) or radiative accelerations, though it includes an {\it ad hoc}
treatment of extra mixing below surface convection zones, when they exist, to
limit the efficiency of diffusion in order to satisfy various empirical
constraints.  All of this is thoroughly discussed by \citet[see their
\S~3]{vbd12} and need not be repeated here.  However, it is worth mentioning
that evolutionary calculations which employ much more sophisticated treatments
of diffusive processes (notably by \citealt{mrr07}) yield ZAHB luminosities that
differ by $\Delta\,M_{\rm bol} \approx 0.038$ mag compared with those that
neglect this physics (also see \citealt{mrr10}).  This is less than the
determination from Victoria models by only $\sim 0.02$ mag.

\begin{table}
\centering
\caption{Basic Stellar Physics Implemented in the Victoria Code} 
\label{tab:t1}
\smallskip
\begin{tabular}{lll}
\hline
\hline
\noalign{\smallskip}
\multispan2 {\hfil Physics \hfil} & Reference  \\
\noalign{\smallskip}
\hline
\noalign{\smallskip}
 Equation of State & & \citet{vsf00} \\
\noalign{\vskip 2pt}
 Opacities  & Radiative (high $T$)         & \citet{ir96}${^a}$  \\
            & Radiative (low $T$)          & \citet{faa05}${^a}$ \\
            & Conductive                   & \citet{cpp07} \\
\noalign{\vskip 2pt}
 Nuclear Reactions${^b}$ & $^3$He($^3$He,$\,2p$)$^4$He        & \citet{jdz98} \\
                        & $^3$He($^4$He,$\,p$)$^7$Be          & \citet{cbc07} \\
                        & $^7$Be($p,\,\gamma$)$^8$B           & \citet{jms03} \\
                        & $^{14}$N($p,\,\gamma$)$^{15}$O      & \citet{mfg08} \\
                        & $^{12}$C($\alpha,\,\gamma$)$^{16}$O & \citet{hfk05} \\
                        & triple $\alpha$                     & \citet{fdb05} \\
\noalign{\vskip 2pt}
 Neutrino Cooling       &  & \citet{ihn96} \\
\noalign{\vskip 2pt}
 Diffusion              &  & \citet{pm91}  \\
\noalign{\smallskip}
\hline
\noalign{\smallskip}
\end{tabular}
\begin{minipage}{1\columnwidth}
$^{a}$~Opacities for stellar interior conditions were generated via the web site
https://opalopacity.llnl.gov; complementary data for the outer layers of stars
were computed using the code described by Ferguson et al. \\
$^{b}$~For reactions not explicitly identified, rates from the NACRE compilation
(\citealt{aar99}) were adopted. 
\phantom{~~~~~~~~~~~~~~}
\end{minipage}
\end{table}

The adoption of alternative equation-of-state formulations (e.g.,
\citealt{rsi96}), nuclear reactions (e.g., the latest JINA {\it Reaclib}
database; \citealt{caf10}), and radiative opacities (e.g., Opacity Project data;
\citealt{bbb05}) can be expected to have some impact on the predicted
properties of ZAHB models, just as \citet{cps21} have found in their study of
the uncertainties of conductive opacities in moderately to strongly degenerate
gases.  Nevertheless, the differences in the resultant stellar models are likely
to be relatively small given that the basic physics of lower mass stars appears
to be quite well understood.  This is suggested, e.g., by the fact that the
physics in Table~\ref{tab:t1} leads to HB-precursor, RGB-tip models with
luminosities that are in excellent agreement with empirical constraints (see
\citealt{van23}), and by comparisons of stellar models based on completely
independent codes (see below).

The top panel of Figure~\ref{fig:f1} illustrates one of the four grids of ZAHB
models that was computed for this study.  It was generated for metallicities
ranging from [Fe/H] $= -2.5$ to $-0.5$, in steps of 0.2 dex, assuming $Y =
0.250$ and [O/Fe] $= +0.6$, with [$m$/Fe] $= +0.4$ for all of the other $\alpha$
elements.  This mixture of the metals is the same as the one that was given the
name ``{\tt a4xO\_p2}" by \citet{vec22}.  An otherwise identical set of ZAHB
models was computed on the assumption of $Y = 0.290$, while the two remaining
grids adopted the same values of [Fe/H] and $Y$ but the ``{\tt a4xO\_p0}" mix,
which has [O/Fe] $=$ [$\alpha$/Fe] $= +0.4$.  With the aid of a suitable
interpolation code (see below), it is possible to obtain individual ZAHB loci
for any values of [Fe/H], $Y$, and [O/Fe] within the aforementioned ranges. 

\begin{figure}
\begin{center}
\includegraphics[width=\columnwidth]{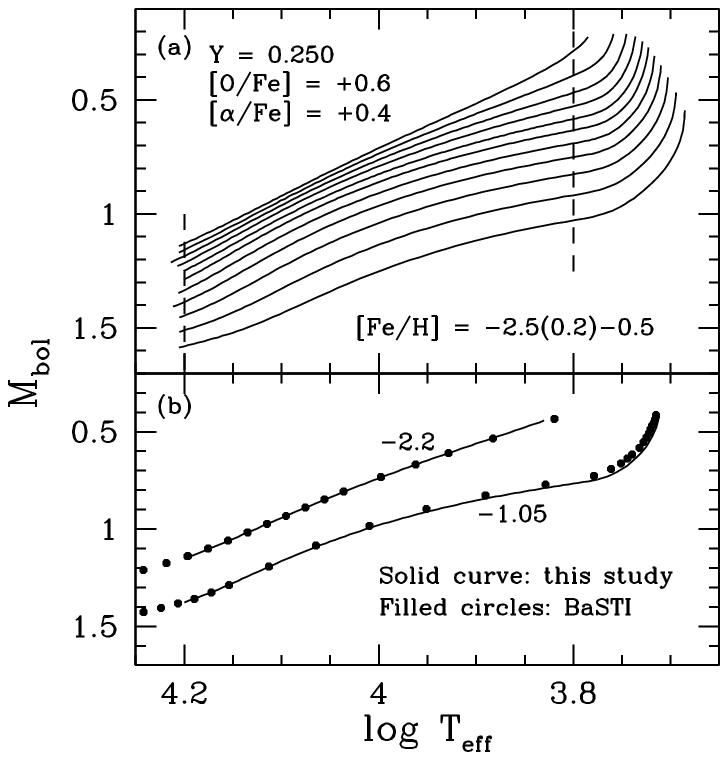}
\caption{Panel (a): grid of ZAHB models for the indicated initial chemical
abundances.  The dashed vertical lines have been plotted simply to illustrate
the approximate temperature range at which interpolations at constant
$\log\teff$ are used to obtain ZAHB loci for any metallicity within the range
$-2.5 \le$ [Fe/H] $\le -0.5$ (see the text).  Panel (b): Comparison of BaSTI
ZAHB models (\citealt{phc21}) for the specified [Fe/H] values with interpolated
ZAHBs from this study for the same metallicities.  Both sets of models assume
$Y \approx 0.250$ and [$m$/Fe] $= +0.4$ for the $\alpha$ elements, including
oxygen.}
\label{fig:f1}
\end{center}
\end{figure}

To generate a sequence of ZAHB models for an adopted initial chemical
composition, a track was computed from a fully convective structure on the
Hayashi line to the tip of the RGB assuming a mass for which the predicted age
at the onset of the He flash is $\approx 12.5$ Gyr.  (Minor variations in the
adopted mass do not affect the resultant ZAHB; see DVKF17,
\citealt{phc21}.)  The chemical abundance profiles from this RGB precursor were
then inserted into a previously converged ZAHB model and the mass-fraction
abundance of $^{12}$C in the core was increased by 0.04 to take into account
the production of carbon that is predicted to occur during the flash (e.g.,
\citealt{swe87}).  The resultant structure was then relaxed over 25 short
timesteps until the age of the model reached 2.5 Myr.  The same procedure was
followed in generating the next model in the sequence except that the mass of
the envelope was reduced by a small amount before that model was relaxed to its
ZAHB location on the H-R diagram.  This process continued until the lowest mass
model in the sequence had a predicted temperature that exceeded $\log\teff =
4.2$.  Several ZAHB models with masses above that of the reference RGB precursor
were also computed so that all ZAHBs include models with masses $\le 0.95 \msol$.
Having all of the ZAHBs run from a common $\teff$ at the hot end to a common
mass at the cool end simplified the writing of a suitable interpolation program.
Each ZAHB typically consists of several dozen stellar models.

Fig.~\ref{fig:f1}b shows that ZAHB models from the latest BaSTI computations
(\citealt{phc21}) are in excellent agreement with those presented in this paper
despite minor differences in the reference solar abundances and in the physics
ingredients (notably the treatment of diffusion) that have been incorporated
into the respective evolutionary codes.  Even the predicted luminosities and
effective temperatures of ZAHB models for the same mass are almost identical.
For instance, the ZAHB sequences for [Fe/H] $= -1.05$ are both terminated at
their cool ends by nearly coincident $0.85 \msol$ models, whereas the highest
masses considered in the ZAHBs for [Fe/H] $= -2.2$ are just slightly offset
from each other, but in the expected given that they have masses of $0.80 \msol$
(BaSTI) and $0.797 \msol$ (this study). It may be recalled (see \citealt{van23})
that the turnoff luminosity versus age relations predicted by BaSTI and
Victoria-Regina isochrones are also essentially indistinguishable when both
assume the same metallicity and  C$+$N$+$O abundance --- though they have
somewhat different capabilities insofar as the fitting of the shapes of observed
CMDs is concerned, which is probably due mostly to differences in the adopted
diffusion physics and bolometric corrections (BCs).

\citet{vdc16} have already demonstrated that the ZAHBs (and isochrones) produced
by the Victoria and MESA (\citealt{pbd11}) codes are in equally fine agreement
when close to the same physics is adopted.  This particular comparison provides
an important test of the reliability of the numerical procedures that are used
in this study (and by \citealt{phc21}) to generate ZAHB models because MESA
follows the evolution of lower mass stars through the He flash, thereby
producing a continuous track between the tip of the RGB and the ZAHB.  Once the
most massive ZAHB model has been created in this way, those for lower masses
are obtained by removing mass from the initial model in small increments, and
then relaxing the resultant structures over many short timesteps.  Such good
agreement of the results from completely independent stellar evolution programs
is truly very satisfying.

As the interpolation of the ZAHB grids is reasonably straightforward, just a few
comments on the adopted procedures should suffice.  At $\log\teff$ values above
some minimum that is common to {\it all} of the ZAHBs (e.g., over the range in
temperature between the two vertical dashed lines in Fig.~\ref{fig:f1}a),
interpolations are carried out at constant values of $\log\teff$, in steps of
0.005 dex, using [Fe/H] as the independent variable, to obtain the masses
and luminosities at the adopted temperatures, for any metallicity of interest
within the range $-2.5 \le$ [Fe/H] $\le -0.5$.  The parts of the ZAHBs that
extend to cooler temperatures (e.g., to the right of the dashed line at
$\log\teff = 3.8$) are divided into 20 equidistant segments, where each
increment in ``distance" is $\delta\,d = [(1.25\,\delta\log L/L_\odot)^2 +
(10\,\delta\log\teff)^2]^{0.5}$.  Once the original ZAHBs in each grid have been
interpolated to give the masses, luminosities, and temperatures at the endpoints
of these segments, these properties are interpolated using \citet{aki70}
splines to yield their values at the selected metallicity.  The result
of this process will be four ZAHBs for the same [Fe/H] value, but for the two
grid values of $Y$ and the two grid values of [O/Fe].  Linear interpolations
are then used, either at constant $\log\teff$ values over the temperature range
that is common to the ZAHBs for different $Y$ and the same oxygen abundances,
or at constant mass at cooler temperatures, to obtain the ZAHBs for the He
abundance of interest.  In the final step, the latter are similarly interpolated,
using [O/Fe] as the independent variable instead of $Y$, to obtain the ZAHB for
the desired chemical abundances ([Fe/H], $Y$, and [O/Fe]).  

\begin{figure}
\begin{center}
\includegraphics[width=\columnwidth]{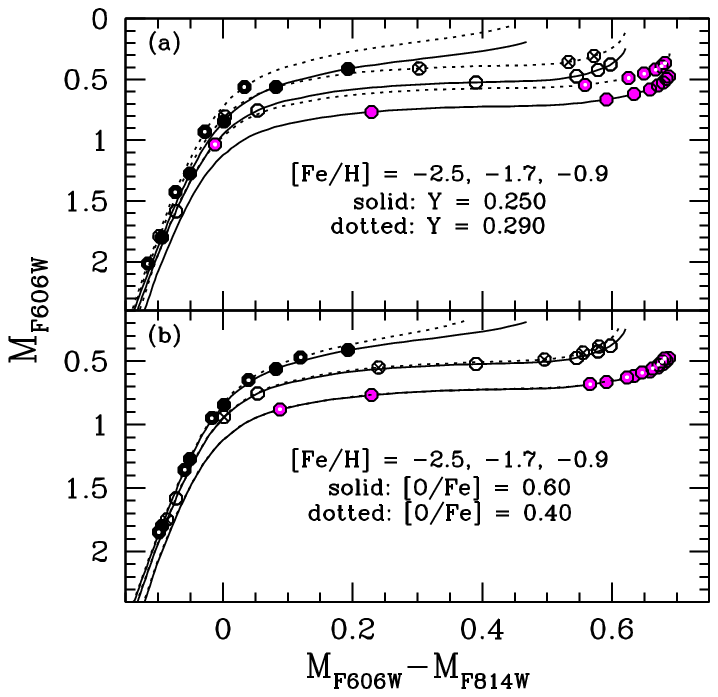}
\caption{Panel (a): comparison of ZAHB loci for the indicated chemical
abundances.  The reddest points that are represented by the various symbols
(i.e., filled circles, open circles, etc.) give the predicted ZAHB locations of
12.5 Gyr stars, assuming that no mass loss occurred during the preceding
evolution.  The bluer points along each ZAHB similarly illustrate, in turn, the
consequences of mass loss amounting to 0.04, 0.08, 0.12, $\ldots \msol$.
Panel (b): as in the top panel, except that the effects of reducing the O
abundance from [O/Fe] $= +0.60$ (solid curves) to $+0.40$ (dotted curves)
are shown.}
\label{fig:f2}
\end{center}
\end{figure}

The solid curves in Figure~\ref{fig:f2} represent the same ZAHBs from
Fig.~\ref{fig:f1}a for [Fe/H] $= -2.5, -1.7,$ and $-0.9$ after they have been
transposed to the $(M_{F606W}-M_{F814W},\,M_{F606W})$-diagram using the BCs
given by \citet{vec22}, which differ only slightly from those supplied by
\citet{cv14} for filters at optical and longer wavelengths.  (In the case
of ultraviolet filters, the 2022 BCs differ significantly from, and improve
upon, earlier transformations; see \citealt{vce22}.)  To illustrate the effects of
varying the metal abundance, as well as the amount of mass loss that would have
occurred prior to reaching the ZAHB, several points have been identified along
each curve.  The reddest of these points give the predicted ZAHB locations of
12.5 Gyr stars if no mass loss is assumed to occur during the preceding
evolution. Thus, for instance, HB stars in 12.5 Gyr GCs with [Fe/H] $= -2.5$
should all have intrinsic $M_{F606W}-M_{F814W}$ colors $\lta 0.19$, whereas
similar HBs in coeval clusters with [Fe/H] $= -1.7$ could be as red as
$(M_{F606W}-M_{F814W})_0 = 0.60$.

\begin{figure}
\begin{center}
\includegraphics[width=\columnwidth]{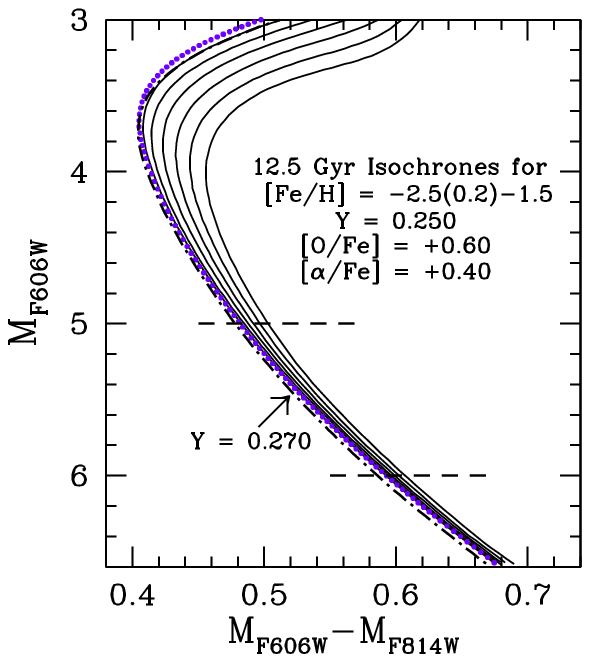}
\caption{Comparison of the upper-MS and TO portions of isochrones for the
indicated ages and chemical abundances.  The dot-dashed curve gives the
location of an isochrone that is otherwise the same as the adjacent solid curve
(for [Fe/H] $= -2.5$) except that it assumes $Y=0.270$ instead of $0.250$.
An isochrone for [Fe/H] $= -2.5$, but with [O/Fe] $= +0.4$ (instead of $+0.6$)
appears as the dotted curve in purple.  The horizontal dashed lines enclose the
magnitude range over which the fiducial sequences of GCs are superimposed (in
\S~\ref{sec:dis}) to constrain the cluster distances and reddenings.}
\label{fig:f3}
\end{center}
\end{figure}

The consequences of mass loss, in increments of $0.04 \msol$, are indicated by
successive bluer points along each ZAHB, in the direction from right to left.
If, e.g., red giants in a 12.5 Gyr GC with [Fe/H] $= -2.5$ were to lose between
0.0 and $0.16 \msol$ during their RGB evolution, they would be expected to
populate the ZAHB over the ranges in magnitude and colour between the brightest
and faintest filled circles in black.  Moreover, 12.5 Gyr clusters with [Fe/H]
$= -1.7$ and $-0.9$ are not predicted to have any blue HB stars unless they lose
$\gta 0.16 \msol$ or $\gta 0.28 \msol$, respectively, before arriving at their
respective ZAHB locations.  It is also apparent from the density of points along
higher metallicity ZAHBs that, once the mass loss has exceeded some threshold,
further incremental increases result in large separations of the respective ZAHB
models (see, e.g., the results for [Fe/H] $= -0.9$).  Thus, most of the HB stars
in metal-rich clusters will necessarily be concentrated in red clumps, while the
HBs of intermediate-metal-poor GCs can be expected to span fairly broad colour
ranges.  The plots of cluster HBs that are presented in \S~\ref{sec:dis}
leave little doubt as to the overriding importance of metallicity and
star-to-star mass-loss variations in determining their distributions as a
function of colour --- something that has been known since the pioneering work
by \citet{roo73}.

It has long been recognized, as well, that helium, [CNO/Fe], and age may play
important roles in determining the morphologies of observed HBs (e.g.,
\citealt{roo73}, \citealt{ldz90}, and the review by \citealt[and references
therein]{cat09}).  Higher $Y$, which has the effect of increasing the
luminosities of HB stars and shifting their distributions to bluer colours, as
shown in Fig.~\ref{fig:f2}a, could well be primarily responsible (see, e.g.,
\citealt{cd05}, DVKF17) for the extended blue HB tails that are
characteristic of many GCs; see \citet{bcd16} for plots of the HB populations
of the GCs that were included in the {\it HST} UV Legacy Survey carried out by
\citet{pmb15}.  However, this complication should not be a concern for the
determination of GC distances from ZAHB models, provided that a significant
fraction of the cluster stars have close to the primordial He abundance; recall
the discussion of this point in \S~\ref{sec:intro}.  The bottom panel of
Fig.~\ref{fig:f2} shows that bluer HBs can also be produced in GCs if they have
reduced O, or more generally, lower C$+$N$+$O abundances, but such variations
clearly have only minimal effects on the ZAHBs themselves.  As already
mentioned, ZAHBs are age-independent, though higher ages imply lower TO masses
and therefore somewhat bluer HB distibutions.  

An important feature of CMDs that are constructed from optical and longer
wavelength observations is that the locations of the blue tails of ZAHBs (at,
e.g., $M_{F606W}-M_{F814W} \lta 0.05$ in Fig.~\ref{fig:f2}) are only weakly
dependent on [Fe/H], $Y$, and [O/Fe], especially at very low metallicities
([Fe/H] $\lta -1.7$).  As a result, fits of very blue HB populations to ZAHB
loci depend almost entirely on their apparent distance moduli and reddenings.
However, because ZAHBs are so steeply sloped at blue colours, even small errors
in the reddenings will translate to large errors in the inferred distance moduli
{\it unless} other constraints are invoked.  Such a constraint is suggested by
Figure~\ref{fig:f3}, which shows that the upper-MS portions of isochrones, when
plotted on the same CMDs that are constructed from $F606W,\,F814W$ observations,
have very little dependence on the assumed chemical abundances; at $M_{F606W} >
5.0$, in particular, the isochrones for [Fe/H] $< -1.9$ are indistinguishable,
while the colour separations of iscochrones for higher metallicities that differ
by 0.2 dex amount to only a few thousandths of a magnitude. 

The differences in the predicted $M_{F606W}-M_{F814W}$ colours at $M_{F606W} =
5.5$ are listed in Table~\ref{tab:t2} for isochrones which are very similar
to those that appear in Fig.~\ref{fig:f3}, except for the assumption of a slight
variation of $Y$ with [Fe/H], as noted therein.  According to these results, the
median MS fiducial sequences of GCs with [Fe/H] values between $-2.5$ and $-2.3$
(and the same values of [$\alpha$/Fe]) are expected to have the same colour to
within 0.0017 mag.  Even in the case of clusters with $-1.7 \le$ [Fe/H] $\le
-1.5$, the median $(M_{F606W}-M_{F814W})_0$ colour at $M_{F606W} \approx 5.5$
should not differ by more than 0.0057 mag.  In fact, cluster-to-cluster
variations in [O/Fe] are inconsequential in this regard (see Fig.~\ref{fig:f3}),
though the colours of MS stars are fairly sensitive to the abundance of helium.
According to computed isochrones, enhanced helium by $\Delta\,Y = 0.02$ results
in a bluer MS by $\approx 0.005$ mag, nearly independently of [Fe/H].  However,
this is still a relatively minor complication given that {\it median} He
abundance enhancements due to self-enrichment appear to be at the level of
$\Delta\,Y \lta 0.01$ for most GCs (see \citealt[their Table 4]{mmr18}).  At
$\sim 1.5$ mag below the TO, the CMD locations of isochrones are obviously
independent of age.

\begin{table}
\centering
\caption{Colour Offsets at $M_{F606W} = 5.5$ between Isochrones for different
values of [Fe/H] and $Y^{a}$} 
\label{tab:t2}
\smallskip
\begin{tabular}{ccccccc}
\hline
\hline
\noalign{\smallskip}
 [Fe/H]$_1$ & $Y_1$ & [Fe/H]$_2$ & $Y_2$ & $\Delta(M_{F606W}-M_{F814W})$ \\
\noalign{\smallskip}
\hline
\noalign{\smallskip}
 $-2.5$ & 0.250 & $-2.3$ & 0.250 & 0.0017 \\
 $-2.3$ & 0.250 & $-2.1$ & 0.250 & 0.0024 \\
 $-2.1$ & 0.250 & $-1.9$ & 0.250 & 0.0033 \\
 $-1.9$ & 0.250 & $-1.7$ & 0.251 & 0.0044 \\
 $-1.7$ & 0.251 & $-1.5$ & 0.251 & 0.0057 \\
 $-1.5$ & 0.251 & $-1.3$ & 0.252 & 0.0074 \\
\noalign{\smallskip}
\hline
\noalign{\smallskip}
\end{tabular}
\begin{minipage}{1\columnwidth}
$^{a}$~The helium abundance is assumed to increase slowly with increasing
[Fe/H] due to Galactic chemical evolution according to $\Delta\,Y/\Delta\,Z =
1.0$, resulting in, e.g., $Y = 0.257$ at [Fe/H] $= -0.70$, which is the minimum
value of $Y$ that was adopted in the simulations of the 47 Tuc HB by DVKF17.
\phantom{~~~~~~~~~~~~~~~}
\end{minipage}
\end{table}

Despite sharing many attributes (steep slopes, little or no sensitivity to age 
or metallicity), there is one crucial difference between the blue tails of ZAHBs
and the portions of isochrones that are relevant to upper-MS stars; namely, the
blue tails of ZAHBs have negative slopes while isochrones have positive slopes.
Consequently, if a higher reddening were adopted, the fitting of an observed HB
to the same ZAHB would yield a smaller distance modulus, while the fitting (at
$M_{F606W} > 5.0$) of the MS stars in the same GC to the MS fiducial sequences
of other GCs with similar metallicities would result in a larger distance
modulus.  In principle, therefore, consistent fits of both the HB and MS
populations in GCs are possible only for certain specific values of $E(B-V)$
and $(m-M)_V$.  

The main point to be drawn from Fig.~\ref{fig:f3} is that, when the
HB populations of clusters of very low metallicities are fitted to ZAHB loci,
the inferred reddenings and distance moduli must be such that their MS fiducial
sequences superimpose one another.  Importantly, the approach used here to
determine these properties have very little dependence on the cluster
metallicities and their uncertainties, and no dependence whatsoever on the
cluster ages.  The next section describes in more detail the methods that have
been used to determine the apparent distance moduli of GCs.

\section{Caveats and Methodology}
\label{sec:fits}

When VBLC13 compared their ZAHBs, which are very similar to those computed for
this study, with the {\it HST} Advanced Camera for Surveys (ACS) photometry
obtained by \citet[hereafter SBC07]{sbc07}, they found that the models matched
the lower bounds of the observed distributions of HB stars quite well when
reddenings from the \citet[hereafter SFD98]{sfd98} dust maps and metallicities
given by \citet[hereafter CBG09]{cbg09} were adopted.  In particular, there were
no obvious discrepancies between the predicted and observed colours along the
bluest HBs.  However, it was necessary to apply a blueward adjustment to the
isochrones that were fitted to the upper MS and TO observations by
$\delta(M_{F606W}-M_{F814W}) \sim 0.015$--0.03 mag in order to reproduce the
TO colours.  Although the main cause of such discrepancies was not known at the
time, the new reduction and calibration of the same data by NLP18 has
largely eliminated this problem (see \citealt{vce22}).

Since this recalibration also impacts the colours of HB stars, ZAHB models are
no longer able to reproduce their CMD locations quite as well.  To obtain
fits of observed HB populations to ZAHB loci that are similar to those reported
by VBLC13, it is necessary to adjust the predicted $M_{F606W}-M_{F814W}$ colours
by $\approx +0.012$ mag.  Unfortunately, this offset varies from cluster to
cluster by a few to several thousandths of a magnitude, even when considering
GCs that have the same metallicities to within 0.1 dex.  Similar differences
exist between the observed photometry of upper-MS stars given by SBC07, on the
one hand, and by NLP18, on the other; e.g., the median TO colours of
NGC$\,$5053, M$\,$92, and M$\,$68 that are derived from the 2007 and 2018
investigations differ by 0.013, 0.015, and 0.025 mag, respectively.  Although
such differences are within the photometric uncertainties tabulated by NLP18,
they present a major problem for any study, such as the present one, that
requires highly homogeneous photometry.  It is clear from the discussion of
Figs.~\ref{fig:f2} and~\ref{fig:f3} and Table~\ref{tab:t2} that
{\it differences} in the observed $(M_{F606W}-M_{F814W})_0$ colours as small as
$\sim 0.002$ mag matter; i.e., the adopted photometry must be exceedingly
accurate in a differential sense.

Considerable time and effort was expended in examining both the original ACS
$F606W,\,F814W$ observations provided by SBC07 and the recalibrated
versions of the same data given by NLP18.  For most GCs, the HB photometry from
the latter source could be fitted quite well to the ZAHB models (aside from the
small zero-point offset mentioned above) and it was usually, but not always,
possible to obtain satisfactory superpositions of the MS fiducial sequences of
clusters of similar [Fe/H] on the assumption of the best estimates of the
foregound reddenings.  However, satisfactory consistency could not be
found for several clusters (including, e.g., M$\,$68, M$\,$2, NGC$\,$6752).
Because no such difficulties were encountered in similar analyses
of the original ACS photometry by SBC07, the decision was made to adopt these
data in this investigation.  Interestingly, one consequence of this choice is
that good fits of the observed HB populations to ZAHB models could be obtained
on the assumption of the reddenings from dust maps, or very close to them.  Both
of these findings indicated a clear preference for the SBC07 observations, which
appear to be very homogeneous.\footnote{As discussed by
VBLC13, MARCS colour--$\teff$ relations had to be supplemented by BCs based on
Castelli-Kurucz model atmospheres at $\teff > 8000$~K, given that 8000~K is the
maximum temperature for which MARCS model atmospheres and synthetic spectra were
computed (see \citealt{gee08}).  The BCs for higher temperatures were obtained
from S.~Cassisi (2012, priv.~comm.) for the ACS and Johnson-Cousins-2MASS
photometric systems.  As there were significant, albeit small, differences
between the MARCS and Castelli-Kurucz BCs at 8000~K, the Castelli-Kurucz BCs
for higher temperatures were adjusted by amounts that ranged from $-0.018$ to
$+0.012$ mag, depending mainly on the metallicity, to try to improve the
continuity of the two data sets.  Although this bridging of the BCs from the two
different souces was carried out independently of any comparisons of ZAHB models
with observed HBs, the adopted BCs for hot ZAHB models appear to be on precisely 
the same photometric system as the ACS observations that were obtained by SBC07.} 

In this investigation, GCs have been separated into a number of different
metallicity bins.  For those clusters with the bluest HBs, initial estimates of
$(m-M)_V$ were obtained by fitting their HB populations to suitable ZAHB models,
on the assumption of reddenings from dust maps.  As reported by \citet[hereafter
SF11]{sf11}, their reddening maps yield $E(B-V)$ values that are $\approx 86$\%
of those given by the earlier SFD98 maps.  However, roughly half of the
reduction can be attributed to differences in the spectral energy distributions
of the stars that are used to derive the extinction law.  As a result, the
current best estimate of the {\it nominal} reddening of a given cluster ---
i.e., the $E(B-V)$ value that applies to early-type stars, which is the usual
convention for the majority of reddening determinations, including those by
SFD98 --- will be close to $E(B-V)_{\rm SFD98} + 0.5\,[E(B-V)_{\rm SF11} -
E(B-V)_{\rm SFD98}]$, which is equivalent to the average $E(B-V)$ value from
the two reddening 
maps.\footnote{The web site https://irsa.ipac.caltech.edu/applications/DUST/
provides the $E(B-V)$ values and the associated $1\,\sigma$ uncertainties that
were used in this study.}  With this choice, the actual colour excesses that
apply to the upper-MS populations in GCs can be calculated using the values
of $R_\eta$ given by \citet[their Table A1]{cv14}, where $R_\eta = A_\eta/E(B-V)$
and $A_\eta$ is the extinction in the $\eta$ bandpass.  Hence, for filters
$\eta$ and $\zeta$, $E(\zeta-\eta) =  (R_\zeta-R_\eta)\,E(B-V)$ and $(m-M)_\eta
= (m-M)_V +(R_\eta-R_V)\,E(B-V)$, where $E(B-V)$ is the nominal reddening.

\begin{figure}
\begin{center}
\includegraphics[width=\columnwidth]{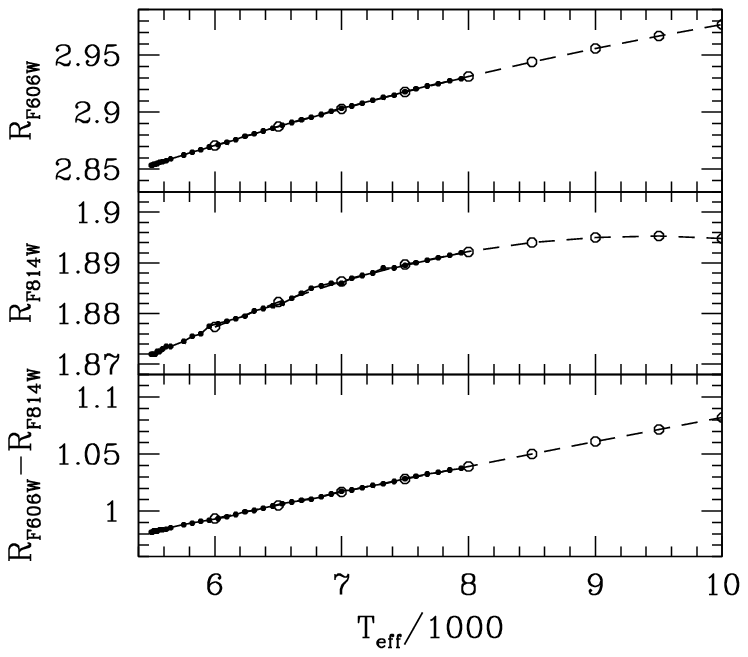}
\caption{Variations of $R_{F606W}$ (upper panel), $R_{F814W}$ (middle panel)
and $R_{F606W}-R_{F814W}$ (lower panel) as a function of $\teff$.  The small
filled circles are based on BCs for $E(B-V) = 0.0$ and 0.20 that have been
derived by interpolations in the tables given by \citet{cv14} at the
temperatures and gravities along a ZAHB for $Y=0.25$ and [Fe/H] $= -2.0$.
These results were fitted by quadratic equations, which were then evaluated at 
the temperatures that are indicated by the open circles.  The extrapolations
to temperatures greater than 8000~K are represented by the dashed curves.} 
\label{fig:f4}
\end{center}
\end{figure}

Because the reddening produced by a given amount of dust is larger for stars
of earlier spectral types, it is important to use appropriate values of $R_\eta$
when dealing with blue HBs.  Unfortunately, as already noted, the BCs provided
by \citet{cv14} to take reddenings up to $E(B-V) = 0.70$ into account, are 
limited to stars with $\teff \le 8000$~K.  However, Figure~\ref{fig:f4} shows
that the dependencies of $R_{F606W}$, $R_{F814W}$, and $R_{F606W}-R_{F814W}$ on
$\teff$ are sufficiently weak and well defined that the relations which are
obtained from ZAHB models with $\teff < 8000$~K can be reliably extrapolated to
higher temperatures --- specifically to $\teff = 10,000$~K, which is close to
the temperature at which ZAHB loci on optical CMDs make a transition from steep
blue tails to nearly horizontal morphologies.  Based on the results shown in
Fig.~\ref{fig:f4}, the extinctions and the reddenings of observed HB stars
located near the tops of blue tails have been calculated assuming $R_{F606W} =
2.977$ and $R_{F814W} = 1.895$.  The robustness of these determinations is
indicated by the fact that the differences in the extrapolated values of
$R_{F606W}$ and $R_{F814W}$ are in excellent agreement with the extrapolated
values of $R_{F606W}-R_{F814W}$ at $\teff > 8000$~K.  (Since $F336W,\,F438W$
observations also play an important role in this investigation, the $R_\eta$
values for these filters have been similarly extrapolated, resulting in
$R_{F336W} = 5.148$ and $R_{F438W} = 4.147$ at $\teff = 10,000$~K.)   
  
Once initial estimates of $(m-M)_V$ have been derived for the most
metal-deficient group of clusters from overlays of their blue HBs onto
appropriate ZAHB models, assuming dust-map reddenings, their upper-MS fiducial
sequences are intercompared to determine whether they superimpose one another
sufficiently well to be consistent with the theoretical results that are
presented in Fig.~\ref{fig:f3}.  If this constraint is not satisfied, the
initial estimates of $(m-M)_V$ and/or $E(B-V)$ are adjusted in order to obtain
the expected consistency.  Fortunately, the second step in this two-step
process usually yields just a minor refinement of the cluster properties (see
the next section), which indicates that the ZAHB models do quite a good job of
matching blue HBs without having to take into account the MS constraints.  In
fact, consistent results for the HB and MS populations of most of the GCs
considered in this study could be obtained on the assumption of the reddenings
from dust maps that are within the ranges implied by their $1\,\sigma$
uncertainties, which are typically $\lta 0.002$ mag for nearly unreddened
systems.
 
\begin{figure*}
\begin{center}
\includegraphics[width=0.96\textwidth]{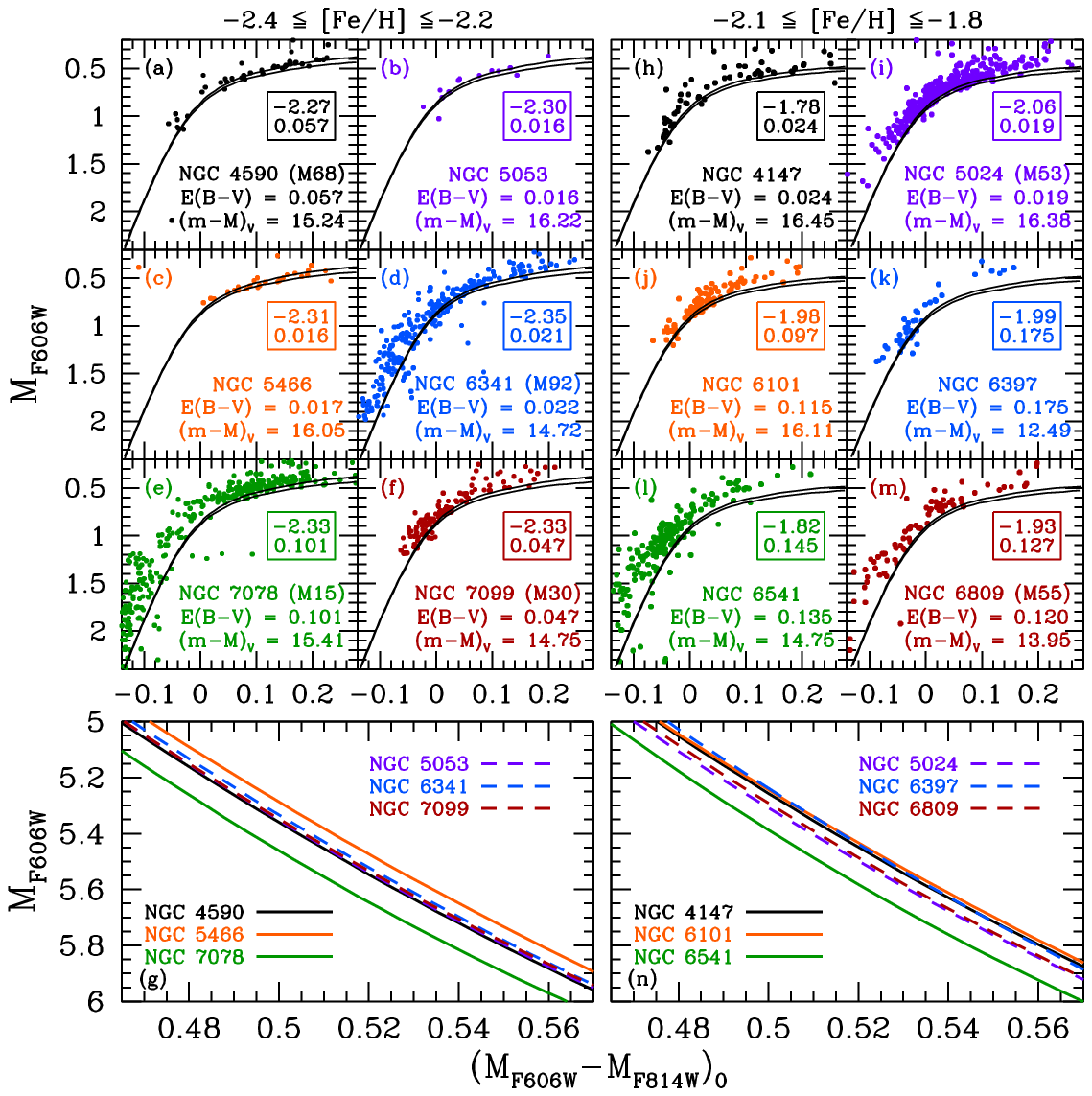}
\caption{Panels (a)-(f) and (h-(i): the adopted fits of the HB populations
in several GCs to ZAHB loci (solid curves) that have been generated for [Fe/H]
values corresponding to the minimum and maximum metallicities that define each
of two bins (identified at the top of the plot).  These fits and the
superpositions of the cluster MS fiducials in panels (g) and (n) are obtained
when the indicated values of $E(B-V)$ and $(m-M)_V$ are assumed.  The small
boxes in each of the smaller panels give, in turn, the cluster metallicities
from the spectroscopic survey by CBG09 and the best estimates of the $E(B-V)$
values from dust maps.}
\label{fig:f5}
\end{center}
\end{figure*}

For those GCs with red HB populations, the derived values of $(m-M)_V$ that
are obtained from the superpositions of the observed HB stars onto appropriate
ZAHB models have very little dependence on the adopted reddenings (because ZAHB
loci and observed HBs are nearly horizontal) or on the assumed metallicities.
Indeed, comparisons of the cluster CMDs in the vicinity of the TO are not needed
to constrain the fits of cluster HBs to ZAHB models when the observed HBs span
broad colour ranges, though they do provide a useful check of the relative
cluster metallicities given by CBG09.  The most challenging clusters to analyze
are those with the bluest HBs, such as M$\,$13 and NGC$\,$6752.  However, it
turns out that $(M_{F336W}-M_{F606W})_0$ colours provide valuable constraints
on the relative reddenings and apparent distance moduli of such systems.   This
will be demonstrated in the next section, where many intercomparisons of
UV-optical CMDs are presented and discussed.

\section{ZAHB-based Apparent Distance Moduli}
\label{sec:dis}

\subsection{GCs that have $-2.3 \lta$ [Fe/H] $\lta -1.8$}
\label{subsec:lowz}

Figure~\ref{fig:f5} illustrates the results that are obtained for GCs in the
two lowest metallicity bins when (i) their HBs are fitted to ZAHBs for [Fe/H] $=
-2.4$ and $-2.2$ (in panels a--f) or those for $-2.1$ and $-1.8$ (in panels h--m),
and (ii) their MS fiducials superimpose one another quite well in the magnitude
range $5.0 \le M_{F606W} \le 6.0$; see panels (g) and (n).  Given the rather
small separations of ZAHBs arising from a 0.2 dex difference in the assumed
[Fe/H] values, uncertainties in the cluster metallicities amounting to $\pm 0.1$
dex clearly have relatively minor consequences for the apparent distance moduli
that are derived from ZAHB models.  Note that adjacent tick marks in the bottom
panels correspond to a colour difference of only 0.005 mag, which is comparable
to or greater than the widths of the bands that encompass the upper-MS fiducial
sequences of most of the GCs in each bin.  To facilitate examinations of this
and similar plots, the average reddenings from the SF11 and SFD98 dust maps and
the cluster metallicities from CBG09 are specified in small boxes in each of the
smaller panels.  According to Carretta et al., for instance, the most
metal-deficient clusters that have been considered (M$\,$68, NGC$\,$5053,
$\ldots$) all have the same [Fe/H] values to within 0.06 dex.

When considering GCs that have predominately blue HBs, the clusters in each
metallicity bin that have the lowest reddenings and/or the reddest HB
populations are of particular importance because they provide the most robust
fits to the ZAHB models and the most secure definition of the CMD location of
the MS fiducial that should be common (or nearly so) to all of the clusters.
Fortunately, most of the GCs with [Fe/H] $\approx -2.3$ have low reddenings,
and it is possible to obtain satisfactory fits of the cluster HBs to ZAHB
models, simultaneously with a near coincidence of their MS fiducials, if dust
map reddenings, to within their $1\,\sigma$ uncertainties are adopted.
However, it proved to be more difficult to obtain such consistency in the case
of NGC$\,$5466.  Even if its HB stars are fitted to a ZAHB for [Fe/H] $\approx
-2.20$ (see panel c), instead of one for the metallicity given by CBG09, its MS
is offset from those of the other GCs that have similar metallicities by about
0.005 mag (see panel g).  Although this discrepancy could be reduced if a
higher reddening were adopted, it is also possible that NGC$\,$5466 has a
metallicity closer to [Fe/H] $= -2.0$ than to $-2.3$, as found in the recent
spectroscopic analysis by \citet{lvs15}.  Regardless, the adopted ZAHB-based
apparent distance modulus, $(m-M)_V = 16.03$, would not change by more than
$\approx 0.02$--0.03 mag in view of the relatively weak sensitivity of the ZAHB
models to [Fe/H].

\begin{figure*}
\begin{center}
\includegraphics[width=0.96\textwidth]{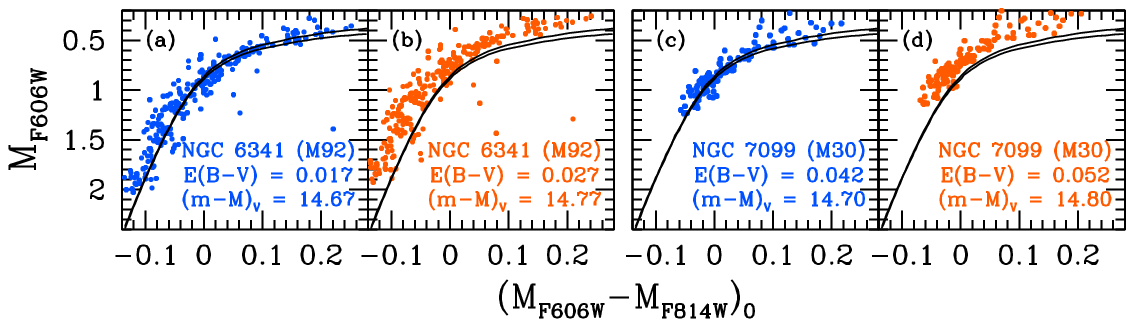}
\caption{Overlays of the same ZAHBs for [Fe/H] $= -2.4$ and $-2.2$ that appear
in the previous figure onto the HB stars in M$\,$92 and M$\,$30 on the
assumption of the indicated values of $E(B-V)$ and $(m-M)_V$.  The CMDs plotted
in orange and black assume reduced or increased reddenings, respectively, by
0.005 mag.  If these reddenings are adopted, the superpositions of the cluster
MS fiducials are comparable to those shown in Fig.~\ref{fig:f5} only if the
specified values of $(m-M)_V$ are adopted.  The resultant cluster parameters
imply the fits of the cluster HBs to the ZAHB models that are illustrated.} 
\label{fig:f6}
\end{center}
\end{figure*}

M$\,$92 also has a very low reddening, and if the best estimate of $E(B-V)$ from 
dust maps is adopted, its MS stars are found to be just slightly redder than 
those in M$\,$68, NGC$\,$5053, and M$\,$30 despite having a similar, or slightly
lower, metal abundance.  Indeed, since M$\,$92 appears to contain stars with
enhanced He abundances (see \citealt{zmm23}), it probably also has a higher mean
value of $Y$ than the other three GCs, which could explain why its HB spans a
wider range in colour and extends to bluer colours.  Higher $Y$ should result
in a bluer MS, though only by two or three thousandths of a magnitude if the
{\it mean} enhancement is as large as $\Delta\,Y \approx 0.01$ (recall
Fig.~\ref{fig:f3}).  While this discrepancy may simply be an indication
that the reddening of M$\,$92 from dust maps is too low by $\gta 0.003$ mag, an
intrinsically slightly redder-than-expected MS could arise if the core 
H-burning stars rotate sufficiently rapidly; see \S~\ref{sec:second} for some
discussion of this possibility. 

The requirement that the MS fiducial seqences of GCs superimpose, or are offset
by a fixed amount from, one another places rather tight constraints on the
ZAHB-based apparent distance moduli.  Suppose, for instance, that a higher or
lower reddening by 0.005 mag were adopted for M$\,$92.  In order to recover the
same CMD location that its MS had on the assumption of $E(B-V) = 0.022$, it
would be necessary to increase or decrease, respectively, the apparent distance
modulus by approximately 0.05 mag.  However, doing so would cause a significant
discrepancy between the cluster HB stars and the appropriate ZAHB models, as
illustrated in the two left-hand panels of Figure~\ref{fig:f6}.  Clearly, there
are too many stars below the ZAHBs in panel (a), whereas the HB stars are
somewhat too bright relative to the ZAHBs in panel (b).  This occurs because the
change in the distance modulus that would be needed to compensate for a change
in the reddening is in the opposite sense for the fitting of the HB stars to a
ZAHB than for the recovery of the original MS location.  To improve upon the
results that are shown in panels (a) and (b), one would need to adopt
intermediate values of $E(B-V)$ and $(m-M)_V$, as in Fig.~\ref{fig:f5}d.  Thus,
the MS constraint limits the uncertainty of the derived distance modulus to
$\sim \pm 0.03$--0.04 mag.  (The right-hand panels similarly show the 
implications for the fitting of the HB stars in M$\,$30 to ZAHB models when the
cluster parameters are derived solely from the MS fits.)

\begin{figure}
\begin{center}
\includegraphics[width=0.96\columnwidth]{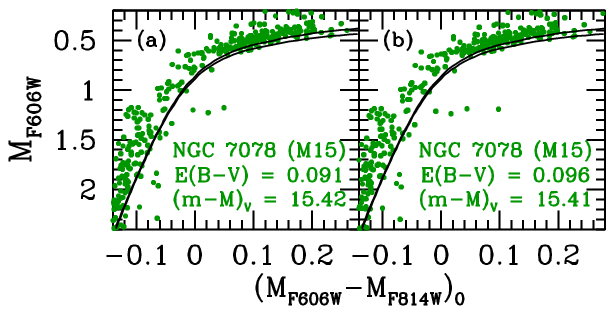}
\caption{As in Fig.~\ref{fig:f5}e, except for small differences in the
adopted cluster parameters (as indicated).}
\label{fig:f7}
\end{center}
\end{figure}

Of all of the GCs in this group, M$\,$15 is the most likely one to have large
star-to-star He abundance variations given that such variations would provide a
natural explanation for its very extended blue HB.  This is suggested by the
simulations of the similarly extended M$\,$13 HB carried out by DVKF17, who were
able to reproduce the observed morphology if M$\,$13 has a spread in $Y$
amounting to $\Delta\,Y \sim 0.08$ and an average He abundance close to $Y =
0.285$.  If M$\,$15 is similar to M$\,$13 in this respect, its median MS
fiducial should be significantly bluer at a given absolute magnitude than those
of the other GCs with similar metallicities, and indeed this is found to be the
case if $E(B-V) = 0.101$ (from dust maps) is adopted for M$\,$15 (see
Fig.~\ref{fig:f5}g); the $1\,\sigma$ uncertainty of this determination is
0.003 mag.  In fact, a moderately large variation in $Y$ is suggested by
so-called ``chromosome maps" (\citealt{mmr18})\footnote{There are some concerns
about the reliability of the He abundance spreads that are inferred from such
maps.  Whereas chromosome maps imply $\Delta\,Y \sim 0.10$ for the 1G stars
in M$\,$3 (see \citealt{mmr18}), \citet{tdc19} have presented compelling 
arguments that the actual variation within the 1G population is negligible.
These discrepant findings have not yet been satisfactorily explained.  On the
other hand, chromosome maps indicate that the average spread in $Y$ between the
1G and 2G populations in M$\,$3 is $0.016 \pm 0.005$ (\citealt{mmr18}),
which is only slightly higher than the results from synthesized HBs (DVKF17).}
and by the analysis of the properties of the many RR Lyrae variables in this
cluster by \citet{vdc16}.

M$\,$15 may have a lower reddening, which would reduce, possibly even eliminate,
the separation of its MS relative to those of other GCs of similar [Fe/H], but
as shown in Figure~\ref{fig:f7}, this would alter the derived value of $(m-M)_V$
by only 0.01--0.02 mag.  The reason why the derived distance modulus is not
particularly dependent on $E(B-V)$ is the presence of HB stars with
$(M_{F606W}-M_{F814W})_0 \gta 0.05$, where ZAHBs have relatively shallow slopes.
Even in the small versions of the plots that are shown, it is clear that the
ZAHB provide identical fits to the faintest stars with red colours as in
Fig.~\ref{fig:f5}e.  Indeed, $(m-M)_V = 15.41$ appears to be quite a robust
determination of the distance modulus of M$\,$15.

\begin{figure*}
\begin{center}
\includegraphics[width=0.96\textwidth]{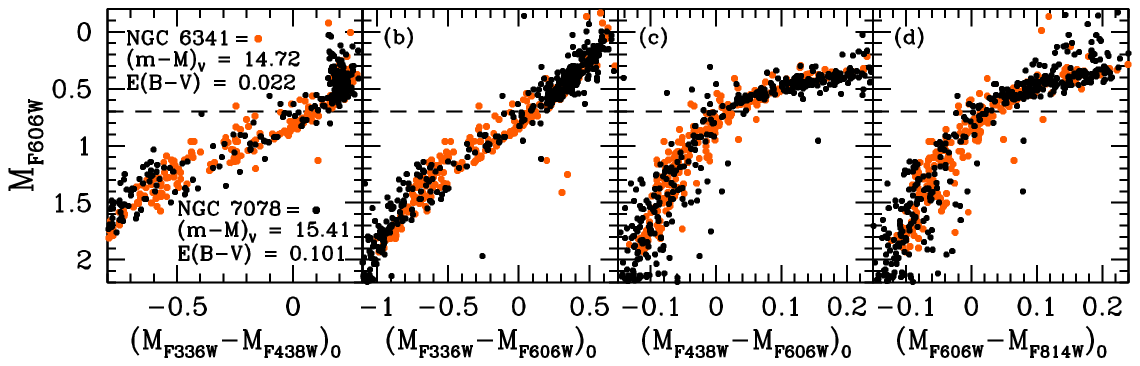}
\caption{Superposition on 4 different CMDs of the HB stars (black filled
circles) in NGC$\,$7078 (M$\,$15) onto the HB population (orange filled circles)
of NGC$\,$6341 (M$\,$92).  The adopted values of $E(B-V)$ and $(m-M)_V$, which
are specified in the left-hand panel, are identical to the values that were
adopted in the construction of Figs.~\ref{fig:f5}d.  The horizontal dashed line
gives the approximate location at which a transition is made between the steep
blue tails and the more horizontal parts of ZAHBs when plotted on optical CMDs
(panels c and d).}
\label{fig:f8}
\end{center}
\end{figure*}

For the 6 GCs considered in this investigation that have $-2.1 \lta$ [Fe/H]
$\lta -1.8$ (see Fig.~\ref{fig:f5}), the assumption of dust-map reddenings (to
within $1\,\sigma$) results in good fits of the HB stars in all of the
clusters except NGC$\,$6101 and M$\,$55 to the relevant ZAHBs and, with the
exception of NGC$\,$6541, quite satisfactory superpositions of their MS
fiducial sequences (see panel n).  While the M$\,$53 MS is offset to the blue
by about 0.005 mag, it is also the most metal-poor cluster in this group.  An
examination of panels (g) and (n) reveals that its MS is located roughly midway
between the main sequences of NGC$\,$4147 ([Fe/H] $= -1.78$) and 
NGC$\,$5053 ([Fe/H] $= -2.30$), which is reasonably consistent with an [Fe/H] of
$-2.06$.  The HBs of NGC$\,$4147, M$\,$53, and NGC$\,$6397 span relatively small
colour ranges, which suggests that they have similar He abundances with little
or no spread in $Y$.  \citet{mmp12} found that NGC$\,$6397 has a split MS, from
which they concluded that 30\% of the stars have close to the primordial He
abundance and the rest have slightly higher $Y$ by $\sim 0.01$.  The mean He
abundances of NGC$\,$4147, M$\,$53, and NGC$\,$6397 are probably not very
different.

A satisfactory fit of the NGC$\,$6101 HB to ZAHB models could not be obtained
on the assumption of $E(B-V) = 0.097$ (from dust maps), which has a $1\,\sigma$
uncertainty of 0.005 mag.  Such a low reddening would result in the cluster MS
being {\it redder} than the location of the orange curve in Fig.~\ref{fig:f5}n
by more than 0.02 mag.  Note that a lower reddening must be accompanied by an
increased distance modulus in order to fit the cluster HB stars to the ZAHBs,
which exacerabates the discrepancy.)  As there is no obvious way of explaining
such a red MS, especially when the HB morphology of NGC$\,$6101 is so similar
to those of NGC$\,$4147, M$\,$53, and NGC$\,$6397, it would appear that the
reddening from dust maps is too low.  Better overall consistency is obtained if
NGC$\,$6101 has $E(B-V) = 0.115$ and $(m-M)_V = 16.11$, as shown in panels (j)
and (n).  Indeed, the UV-optical CMDs that are discussed in the next section
support these values of the cluster parameters.

Both NGC$\,$6541 and M$\,$55 have extended blue HBs (especially the former; see
\citealt{bcs18}); consequently, it can be anticipated that they have higher
mean He abundances than the other GCs in their respective metallicity groups.
This is, in fact, consistent with the indications from chromosome maps (see
\citealt{mmr18}).  Given the similarity of the blue HBs in NGC$\,$6541 and 
M$\,$15, in particular, it can be expected that both of their median MS
fiducials will be bluer by $\approx 0.01$ mag than they would be {\it if} all of
the member stars had $Y \approx Y_P$.  This is approximately the offset that
is obtained between the MS of NGC$\,$6541 and other GCs that have [Fe/H] $\sim
-2.0$ if NGC$\,$6541 has $E(B-V) = 0.134$, which is less than the dust-map
determination by only $1\,\sigma$. 

A reddening as high as $E(B-V) = 0.145$ (from dust maps) seems unlikely because
this would result in a near superposition of the MS fiducials of NGC$\,$6541 and
M$\,$15.  This is untenable because of the $\sim 0.5$ dex difference in their
metallicities and because the adopted properties of M$\,$15 already imply that
its MS is appreciably bluer than those of, e.g., NGC$\,$5053 and M$\,$30 as a
consequence of helium abundance differences.  On the other hand, if $E(B-V)
\approx 0.125$ (nearly a $2\,\sigma$ reduction) and the same distance
modulus (i.e., $m-M)_V = 14.75$) were adopted, the fit of the cluster HB stars
to the ZAHB models would be somewhat improved over the fit that is shown in
Fig.~\ref{fig:f5}l, at the cost of reducing the offset between the MS fiducial
of NGC$\,$6541 and those of NGC$\,$4147, M$\,$53, and NGC$\,$6397 to $\sim
0.003$ mag.  No colour offset whatsoever would be obtained if NGC$\,$6541 has
$E(B-V) = 0.125$ and $(m-M)_V = 14.77$.  Thus, the assumed He abundance of
NGC$\,$6541 has larger implications for the cluster reddening than for the
apparent distance modulus that is derived from the intercomparisons of the
main sequences of GCs of similar [Fe/H] and the simultaneous fitting of the
observed HB stars to ZAHB models.

\begin{figure*}
\begin{center}
\includegraphics[width=0.96\textwidth]{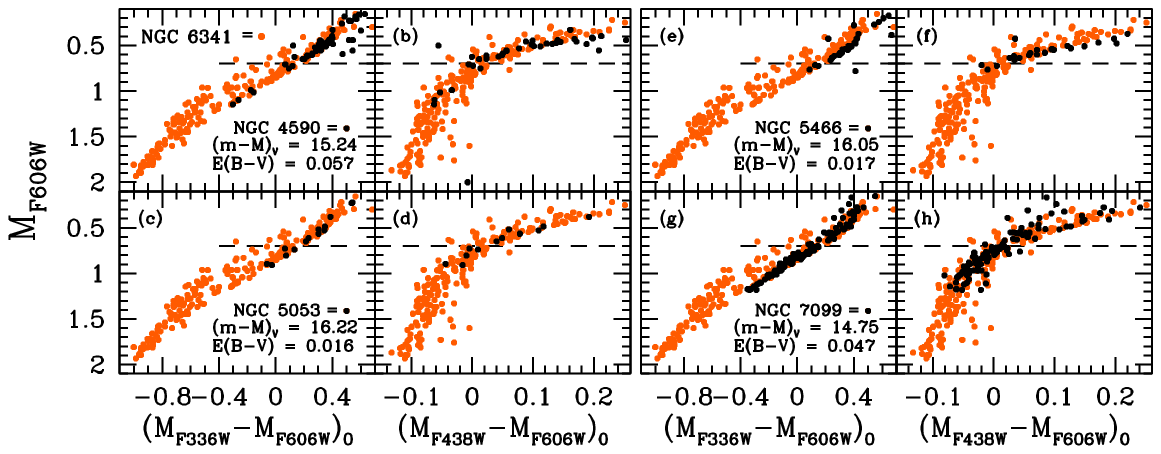}
\caption{Similar to panels (b) and (c) in the previous figure, except that the
HB stars in NGC$\,$4590 (M$\,$68), NGC$\,$5053, NGC$\,$5466, and NGC$\,$7099
(M$\,$30) have been superimposed onto the HB population of NGC$\,$6341
(M$\,$92).  The adopted cluster parameters are the same as in Fig.~\ref{fig:f5}.}
\label{fig:f9}
\end{center}
\end{figure*}

\begin{figure*}
\begin{center}
\includegraphics[width=0.96\textwidth]{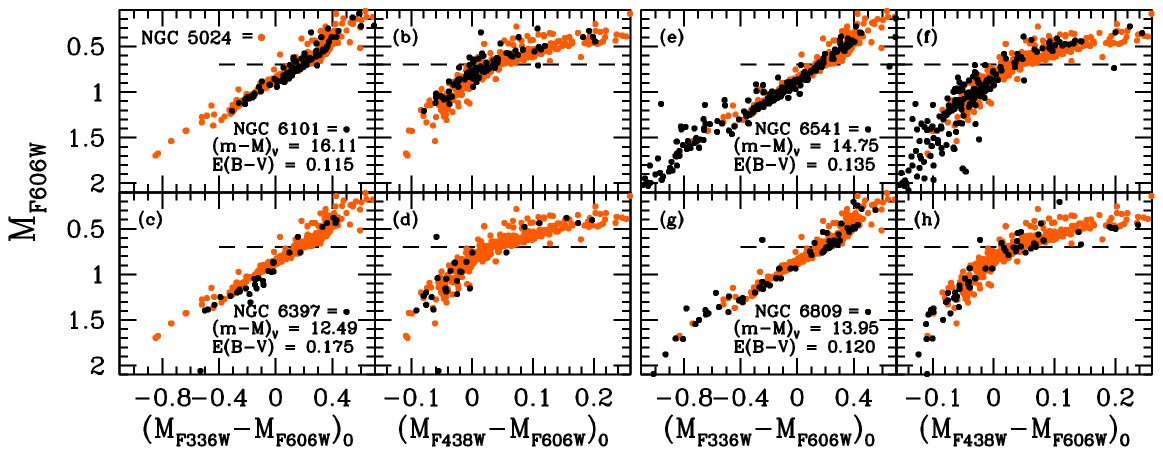}
\caption{Similar to panels (b) and (c) in Fig.~\ref{fig:f8}, except that the
HB stars in NGC$\,$6101, NGC$\,$6397, NGC$\,$6541, and NGC$\,$6809 (M$\,$55)
have been superimposed onto the HB population of NGC$\,$5024 (M$\,$53).  The
adopted cluster parameters are the same as in Fig.~\ref{fig:f5}.}
\label{fig:f10}
\end{center}
\end{figure*}

M$\,$55 has already been the subject of a thorough investigation of its CMD,
its RR Lyrae variables, and its eclipsing binaries by \citet{vd18} using the 
same or very similar stellar models, from which it was concluded that this
cluster has $E(B-V) \approx 0.12$ and $(m-M)_V = 13.95 \pm 0.05$.  Simulations
of the M$\,$55 HB, which were also presented in this paper, indicated that the
He abundance varies by $\Delta\,Y \approx 0.02$, with a median enhancement of
$\lta 0.01$.  It can therefore be expected that the cluster MS should be about
0.005 mag bluer than those of, say, NGC$\,$4147 and NGC$\,$6397, and
indeed, an offset of about this amount is obtained if the cluster parameters
derived by VandenBerg \& Denissenkov are adopted (see Fig.~\ref{fig:f5}n).

\subsubsection{Constraints from HST UV Legacy Photometry}
\label{subsubsec:wfc3}

During the course of this investigation, it was discovered that $F336W,\,F438W$
observations from the {\it HST} UV Legacy Survey provide the means to validate
or to refine the best estimates of the {\it relative} cluster parameters that
are derived from $F606W,\,F814W$ photometry.  This is illustrated in
Figure~\ref{fig:f8}, which superimposes the HB populations of M$\,$92 and 
M$\,$15 on four of the CMDs that can be generated from the Wide Field Camera 3
(WFC3) and ACS observations provided by NLP18.  In contrast with optical
CMDs, in which the morphologies of observed HBs have steep blue tails at
$M_{F606W} \gta 0.7$ (indicated by the horizontal dashed line) and quite shallow
slopes at redder colours, CMDs in which the colours involve $F336W$ magnitudes
have nearly the same slopes over the full colour ranges that have been
considered.\footnote{The synthetic spectra of hot HB stars provided by
\citet[see their Fig.~1]{bcd16} show that the $F336W$ filter samples a part of
the electromagnetic spectrum just on the short wavelength side of the Balmer
discontinuity that contains very few spectral features, especially at low
metallicities.  It would therefore appear that $(M_{F336W}-M_{F606W})_0$ colours
along blue HB tails are primarily a function of $\teff$.} 

Moreover, the first two panels show that observed HB populations of the two
clusters define quite sharp, nearly linear lower boundaries which coincide
almost exactly when the specified reddenings and apparent distance moduli are
adopted.  Indeed, the superpositions of the HBs of M$\,$92 and M$\,$15 in all
four CMDs could hardly be improved upon.  Achieving such consistency across
several different colour planes provides strong support of the relative values
of $E(B-V)$ and $(m-M)_V$ that have been determined.  (Because the CMDs in
panels (a) and (d) are qualitatively very similar to those shown in panels (b)
and (c), respectively, the former have been dropped from consideration in the
additional intercomparisons of NLP18 photometry that are presented and discussed
below.  It is also a considerable space-saving measure to plot two panels for
each cluster, instead of four.) 

Similar comparisons of the HB of M$\,$92 with those of the other GCs with
[Fe/H] $\sim -2.3$ are presented in Figure~\ref{fig:f9}.  In contrast with the
HB populations of M$\,$92 and M$\,$15 (see Fig.~\ref{fig:f8}), the HB stars in
NGC$\,$4590 (M$\,$68), NGC$\,$5053, NGC$\,$5466, and NGC$\,$7099 (M$\,$30)
exhibit almost no scatter in $M_{F606W}$ at a given $(M_{F336W}-M_{F606W})_0$
colour.  A possible, if not the most probable, explanation of the brighter HB
stars in M$\,$92 and M$\,$15 is that they are helium-enhanced stars that have
evolved from ZAHB locations at $(M_{F336W}-M_{F606W})_0 \lta -0.5$ (i.e.,
further down their blue HB tails).
The faintest HB stars in all 6 GCs that superimpose each other along the red
edges of the $(M_{F336W}-M_{F606W})_0,\,M_{F006W}$ diagram when the specified
values of $E(B-V)$ and $(m-M)_V$ are adopted probably have very similar He
abundances, likely close to the primordial abundance.  Clearly, there are no
obvious problems with the adopted cluster parameters (in a relative sense) in
any of the panels that comprise Fig.~\ref{fig:f9}.

The UV-optical CMDs of GCs with $-2.1 \lta$ [Fe/H] $\lta -1.8$, using
NGC$\,$5024 (M$\,$53) as the reference cluster, are similarly intercompared in
Figure~\ref{fig:f10}.  (Plots are not provided for NGC$\,$4147 because this
system does not appear to have been included in the {\it HST} UV Legacy Survey;
see NLP18.)  The HB populations are obviously well centered on one another in
the $(M_{F438W}-M_{F606W})_0,\,M_{F006W}$ diagrams, which serves to confirm
the relative reddenings, in particular.  In fact, the bluest HB stars would be
noticeably offset from their counterparts in NGC$\,$5024 if the adopted
reddenings of NGC$\,$6101, NGC$\,$6397, NGC$\,$6541, and NGC$\,$6809 differed
from the adopted values by more than 0.005 mag, assuming the same distance
moduli.

Encouragingly, even though the HBs of NGC$\,$6541 and NGC$\,$6809 extend to
considerably fainter absolute magnitudes than the HB stars in NGC$\,$5024, it is
quite evident that they superimpose one another remarkably well over the full
ranges in their respective luminosities.  It is interesting that most of the HB
stars in NGC$\,$6397 define a fairly tight sequence in Fig.~\ref{fig:f10}c when
the photometric scatter of the same stars is so large in panel (d).  Although a 
few stars in NGC$\,$6397 fall below the distribution of HB stars in NGC$\,$5024
at $(M_{F336W}-M_{F606W})_0 \approx -0.15$, the majority of them support the
specified cluster parameters.  This serves to confirm the adopted fit of the
cluster HB to appropriate ZAHB models in Fig.~\ref{fig:f5}k.

According to \citet{bv21}, the best estimate of the distance of NGC$\,$6397 from
{\it Gaia} EDR3 parallaxes is $2.458 \pm 0.059$ kpc, which corresponds to
$(m-M)_0 = 11.95 \pm 0.05$.  \citet{bcs18} had previously obtained a true
distance modulus of $11.89 \pm 0.07$ from a trigonometric parallax based on
{\it HST} WFC3 observations.  If $E(B-V) = 0.175$ and $A_V = 3.13\,E(B-V)$
(\citealt{cv14}), the apparent distance modulus derived here (12.49) implies a
true distance modulus of 11.94, which is in excellent agreement with both of
the direct geometric determinations.  Provided that the stellar models
accurately reproduce the properties of observed HB stars, the $1\,\sigma$
uncertainties of ZAHB-based distance moduli for the majority of the lowest
metallicity GCs should not be larger than $\sim \pm 0.03$ mag, especially if
they contain HB stars that populate the transitions of ZAHBs to redder colours;
otherwise it is not possible to achieve consistent simultaneous fits of both
their HB and MS observations.  
This conclusion is supported by the intercomparisons of UV-optical CMDs.

\subsection{GCs that have $-1.7 \lta$ [Fe/H] $\lta -1.5$}
\label{subsec:midz}

\begin{figure*}
\begin{center}
\includegraphics[width=0.96\textwidth]{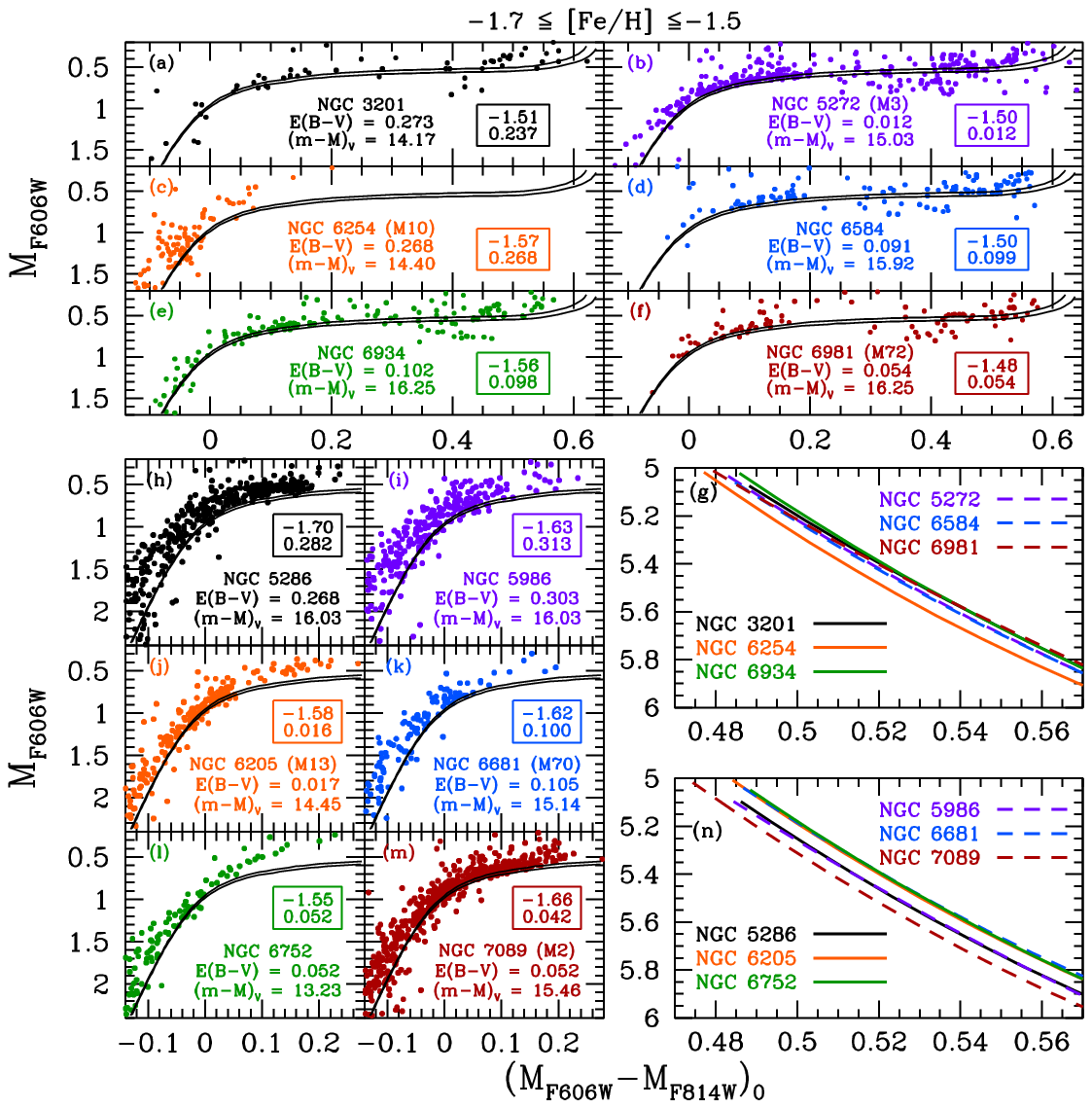}
\caption{As in Fig.~\ref{fig:f5}, except that the HB populations of GCs with
$-1.7 \lta$ [Fe/H] $\lta -1.5$ have been fitted to ZAHBs for the minimum and
maximum metallicities in this range.}
\label{fig:f11}
\end{center}
\end{figure*}

Figure~\ref{fig:f11} presents the fits to ZAHB models of the observed HBs of 
several GCs with metallicities within the range $-1.7 \lta$ [Fe/H] $\lta -1.5$.
Most of the clusters that are considered in the top half of the plot have
moderately large numbers of both blue and red HB stars, spread out over wide
colour ranges, that can be matched to ZAHBs quite easily.\footnote{M$\,$10 was
included in this group mainly to fill the available space, but also because its
HB does not extend as far to the blue as those of the clusters that are
considered in panels (h) to (m).}  Even if their
metal abundances (from CBG09) are in error by $\pm 0.10$ dex, which is possible,
the distance moduli that are derived from such fits would not change by more
than $\pm 0.02$ mag.  Errors in the adopted $E(B-V)$ values or in the predicted
temperatures or colours of the stellar models are of little consequence.
Whether or not the MS fiducials of clusters with the same or similar
metallicities superimpose one another is largely irrelevant for the derived
distance moduli as well, though any inconsistencies that are found would imply
that our understanding of the problematic GCs is deficient in some respect.

According to Table~\ref{tab:t2}, the difference in the $(M_{F606W}-M_{F814W})_0$
colour between isochrones for [Fe/H] $= -1.7$ and $-1.5$ is approximately 0.006
mag at $M_{F606W} = 5.5$.  This is comparable with the width of the band in
panel (g) that encompases the cluster MS fiducial sequences when the specified
values of $E(B-V)$ in panels (a)--(f) are adopted.  These estimates are within
the $1\,\sigma$ uncertainties of the reddenings from dust maps, except in the
case of NGC$\,$3201 and NGC$\,$6584, for which the dust-map determinations are
uncertain by $\pm 0.016$ and $\pm 0.006$ mag, respectively.  To further
investigate this discrepancy, isochrones were fitted to the cluster turnoffs
assuming the adopted reddenings and distance moduli.  The resultant plots
revealed that NGC$\,$3201, M$\,$3, and NGC$\,$6584 are coeval to within $\pm
0.4$ Gyr and, moreover, that they have the same intrinsic TO colours to within
0.003 mag --- as expected given the similarity of their metallicities and ages.
This test provides compelling support for the adopted reddenings of NGC$\,$3201
and NGC$\,$6584 over the $E(B-V)$ values determined from dust maps.  Worth
pointing out is that, when the dust-map reddening is assumed, the MS of M$\,$10
is bluer by about 0.006 mag than those of the other GCs at the same absolute
magnitude (see panel g), which suggests that it may have a high median value of
$Y$.  Although the fitting of the HB stars in M$\,$10 to the ZAHB models is
especially uncertain, UV-optical CMDs (to be discussed shortly) support the
adopted cluster parameters.

The 6 GCs considered in panels (h)--(n) of Fig.~\ref{fig:f11} all have very
extended blue HBs that, in optical CMDs, reach fainter absolute magnitudes
than their TO stars.  They are among the clusters that were given the highest,
``Category 5" ranking by \citet{bcd16} in recognition of their extreme HB
morphologies.  As already mentioned in \S~\ref{sec:models}, a considerable
effort has been made by researchers over the years to determine why the HBs
of some clusters with apparently very similar metallicities, such as M$\,$3 and
M$\,$13, are so different.  At the moment, helium seems to be the leading
contender for the so-called ``second parameter" (after metallicity, which is
the first parameter) insofar as it is possible to simulate observed HBs rather
well if star-to-star {\it spreads} in $Y$ vary from cluster to cluster --- 
though variations in mass loss also appear to play a critical role in the
resolution of this problem; see DVKF17, who also provide quite an extensive
review of this subject.

\begin{figure*}
\begin{center}
\includegraphics[width=0.96\textwidth]{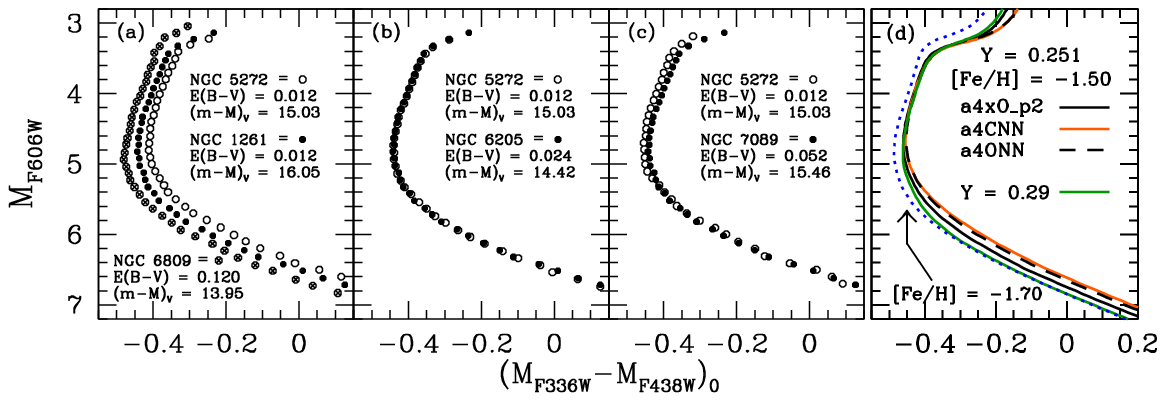}
\caption{Comparisons of the median fiducial sequences for the MS and TO stars
in NGC$\,$5272 (M$\,$3) with those of NGC$\,$6809 (M$\,$55) and NGC$\,$1261 in
panel (a), NGC$\,$6205 (M$\,$13) in panel (b), and NGC$\,$7089 (M$\,$2) in panel
(c), on the assumption of the indicated values of $E(B-V)$ and $(m-M)_V$.
Panel (d) plots the CMD locations of 12.5 Gyr isochrones for the {\tt a4xO\_p2},
{\tt a4CNN}, and {\tt a4ONN} mixtures of the metals, as described in the text,
assuming $Y = 0.251$ and [Fe/H] $= -1.50$.  Otherwise identical isochrones
for the {\tt a4xO\_p2} mix, but for $Y = 0.29$ instead of 0.251 and for [Fe/H]
$= -1.70$ instead of $-1.50$ are represented by the solid green and dotted blue
curves, respectively.}
\label{fig:f12}
\end{center}
\end{figure*}

The HBs of NGC$\,$5286 and NGC$\,$5986 resemble those of M$\,$15 and
NGC$\,$6541 --- all of which are Category 5 clusters.  If, within these systems,
$Y$ varies from $\approx 0.25$ (close to the primordial value) to $\sim
0.30$--0.33, such that the median He abundance is $\sim 0.28$, their upper-MS
stars can be expected to have significantly bluer $(M_{M606W}-M_{F814W})_0$
colours, at the same $M_{F606W}$, than those of GCs with small He abundance
variations.  Relatively large offsets are obtained, in fact, if the reddenings
of NGC$\,$5286 and NGC$\,$5986 are close to dust-map determinations.  On the
other hand, chromosome maps indicate that the average He abundances of the
2G and 1G populations in these systems differ by $\Delta\,Y \lta 0.007$, as
compared with $\Delta\,Y \sim 0.021$ in the case of M$\,$15 (\citealt{mmr18}).
If these findings are trustworthy, the CMD locations of the MS fiducials of
NGC$\,$5286 and NGC$\,$5986 should not be very different from those of
NGC$\,$5272 and NGC$\,$6584, which can be accomplished if $1\,\sigma$ reductions
of the $E(B-V)$ values from dust maps are adopted.  As it turns out, such
reductions result in quite satisfactory fits of the cluster HB stars to ZAHB
models, as shown in Fig.~\ref{fig:f11}h,i.  Note that, whether or not the lower
or higher reddenings are assumed, the derived apparent distance moduli do not
vary by more than 0.02--0.03 mag, especially in the case of NGC$\,$5286 because
its HB extends to fairly red colours.


If the reddenings from dust maps to within their $1\,\sigma$ uncertainties are
adopted for M$\,$13, M$\,$70, NGC$\,$6752, and M$\,$2, the median MS fiducial
sequences of all four GCs would be nearly coincident or intrinsically very
slightly {\it redder} than that of M$\,$3, despite having lower metallicities,
when their HB populations are fitted to ZAHB models.  (If the assumed
reddenings of NGC$\,$5286 and NGC$\,$5986 were reduced by a further 0.007 mag,
their MS fiducials would also superimpose those of the other four GCs.)  Because
of the constraints provided by UV-optical CMDs (see below), the decision was
made to adopt a higher reddening for M$\,$2 by 0.01 mag, in which case the fit
of the cluster HB population to ZAHB models yields $(m-M)_V = 15.46$ (see
Fig.~\ref{fig:f11}m).  A plausible justification for a blueward offset of its
MS (see panel n) is the discovery by \citet{mmp15a} of at least seven chemically
distinct stellar populations in M$\,$2 with He abundance variations as high as
$\Delta\,Y \sim 0.07$.  However, according to chromosome maps (see
\citealt[their Table 4]{mmr18}), there are no statistically significant
differences in the average He abundances between the 2G and 1G stars of M$\,$2,
M$\,$3, or M$\,$13.

On the other hand, HB simulations are able to reproduce the observed HBs of
M$\,$3 and M$\,$13 if the former has a mean He abundance $Y_{\rm avg} = 0.255$
and a spread $\Delta\,Y \approx 0.01$ while the
latter has $Y_{\rm avg} = 0.284$ together with $\Delta\,Y \approx 0.08$ (see
DVKF17).  In this scenario, the MS stars in M$\,$13 and other clusters that have
similar HB morphologies should be intrinsically bluer by $\sim 0.01$ mag at $M_V
\gta 5.0$ than the MS stars in M$\,$3 at the same absolute magnitudes.  However,
to achieve this, it would be necessary to adopt $E(B-V)$ values that are higher
than dust-map estimates by approximately 0.01 mag even though the $1\,\sigma$
uncertainties of these determinations are only 0.001 mag for M$\,$13 and
NGC$\,$6752 (and M$\,$2) as compared with 0.005 mag for M$\,$70.  This seems
rather unlikely, especially as the adopted values of $E(B-V)$ for nearly all of
the clusters considered thus far have been in good agreement with, or less than,
the reddenings from dust maps.  To reiterate this point: {\it it would be quite
odd if M$\,$2, M$\,$13, M$\,$70, and NGC$\,$6752, which all happen to have
exceedingly blue HBs, all have reddenings that exceed dust-map determinations
by such large amounts.}  This anomaly calls into question the possibility that 
the very extended blue HBs that are characteristic of second-parameter clusters 
are mainly due to large star-to-star He abundance variations.

WFC3 UV-optical CMDs {\it for the MS stars} should be able to shed some light on
the possibility that these GCs have high $Y$.  As shown by \citet[see their
Fig.~13]{vce22}, the MS portions of isochrones on
$(M_{F336W}-M_{F438W}),\,M_{F606W}$ diagrams at $M_{F606W} \gta 5.5$
are fairly sensitive functions of
[Fe/H] and $Y$.  They also depend on the abundances of C and N, particularly
when they are typical of the abundances found in CN-strong stars; see 
\citet[their Fig.~10]{vec22}.  Indeed, of the mixtures considered in the latter
study, the {\tt a4CNN} and the {\tt a4ONN} mixes have the strongest effects on
the intrinsic $(M_{F336W}-M_{F438W})_0$ colours along the MS.  Examples of these
findings are illustrated in the right-hand panel of Figure~\ref{fig:f12}.  The
solid black curve represents a 12.5 Gyr isochrone for the indicated values of
$Y$, [Fe/H], and the standard {\tt a4xO\_p2} mix, which assumes scaled-solar
abundances, but with [O/Fe] $= +0.60$ and [$m$/Fe] $= +0.40$ for all other
$\alpha$ elements.  The orange and dashed isochrones have been generated for the
same age, helium abundance, metallicity, and C$+$N$+$O abundance, but whereas
the former assumes the {\tt a4CNN} mixture with reduced C by 0.3 dex and
increased N by 1.13 dex, the latter (i.e., the {\tt a4ONN} mix) assumes lower
abundances of both C and O by 0.8 dex and higher N by 1.48 dex.  Note that the
solid black curve is shifted to the location of the green curve if $Y$ is
increased from a value of 0.251 to 0.29, or to the location of the dotted
blue curve if the metallicity is reduced from [Fe/H] $= -1.50$ to $-1.70$.

Empirical support for the metallicity dependence is provided in panel (a), which
shows that, at a fixed absolute magnitude, the MS of NGC$\,$1261 is appreciably
redder than that of M$\,$3 (NGC$\,$5272), which, in turn, is significantly
redder than the MS of M$\,$55 (NGC$\,$6809.  This is consistent with a variation
in their [Fe/H] values from $\approx -1.3$ (NGC$\,$1261) to $\approx -1.5$
(M$\,$3) to $\approx -1.9$ (M$\,$55).  According to the stellar models (see
panel d), an increased He abundance by $\Delta\,Y \approx 0.04$ has the same
effect on the colours of MS stars at $M_{F606W} \gta 5.8$ as a reduction in the
metallicity by $\Delta\,$[Fe/H] $\approx 0.2$ dex.  It is also apparent in this
panel that some mixtures of C and N can affect $(M_{F336W}-M_{F606W})_0$ colours
in a way that would tend to compensate for the effects on the colours due to
enhanced helium or a reduced metallicity.

Consider M$\,$13 (NGC$\,$6205) and M$\,$2 (NGC$\,$7089), and suppose that the
basic parameters of M$\,$13 are such that its MS is about 0.01 mag bluer than
the MS of M$\,$3 at a given absolute magnitude, similar to what is shown for
M$\,$2.  Since most spectroscopic studies have found that these two clusters
are somewhat more metal deficient than M$\,$3 (e.g., \citealt{ki03}, CBG09),
and if both M$\,$2 and M$\,$13 have higher mean He abundances than M$\,$3 by
$\Delta\,Y \sim 0.035$ to explain their very blue HB morphologies, their main
sequences should be bluer, at $M_{F606W} \gta 5.8$, than the MS of M$\,$3 (see
Fig.~\ref{fig:f12}d). However, the observations indicate otherwise.  The MS
fiducials of M$\,$3 and M$\,$13, in particular, are essentially coincident (see
panel b), which suggests that there is little or no difference in either their
metallicities or their helium contents --- though this conclusion obviously 
depends on the relative distance moduli of the two GCs.  If M$\,$13 has
$(m-M)_V \approx 14.45$ and $E(B-V) = 0.017$ (see Fig.~\ref{fig:f11}j), its MS
would lie slightly above the M$\,$3 MS at redder colours.  This might be
explained, at least in part, by the observation that there are stars in
M$\,$13 with higher C+N abundances than in M$\,$3 (see \citealt{cm05}), even
though both clusters are known to have the same C+N+O abundance to within
measurement errors (also see \citealt{ssb96}).  However, it would be suprising
if the difference in the mean CN strengths of the MS stars in M$\,$3 and M$\,$13
is large enough to cause very much of an offset between the median fiducial
sequences of these two clusters on
the $(M_{F336W}-M_{F606W})_0,\,M_{F606W}$ diagram. 

While the separation between the M$\,$3 and M$\,$2 fiducial sequences at
$M_{F606W} \lta 5.2$ is consistent with about a 0.2 dex difference in [Fe/H]
(see panel c), it does not continue to fainter magnitudes, in contrast with
the M$\,$3--M$\,$55--NGC$\,$1261 intercomparisons in panel (a).  This may be
suggesting that M$\,$2 has a different abundance of C$+$N$+$O (or perhaps a
range in the C$+$N$+$O abundance) and/or different variations in the ratios of
C:N:O than M$\,$3. In fact, \citet[also see \citealt{lpm12}]{yrg14} have found
extreme variations in the observed CN and CH line strengths in spectra of
M$\,$2 stars.  Unfortunately, due to the current lack of model atmospheres
and synthetic spectra that assume wider ranges in both the C$+$N$+$O abundance
and the ratios of C:N:O than those considered by \citet{vec22}, it is not
possible to generate suitable isochrones to further investigate this issue.
Note that, although the differences between the M$\,$2 and M$\,$3 CMDs in panel
(c) could be an indication that the distance modulus of M$\,$2 is too high, any
reduction in the value of $(m-M)_V$ would present a problem for the fitting of
the reddest HB stars in M$\,$2 to ZAHB models (see Fig.~\ref{fig:f11}m).

\begin{figure*}
\begin{center}
\includegraphics[width=0.96\textwidth]{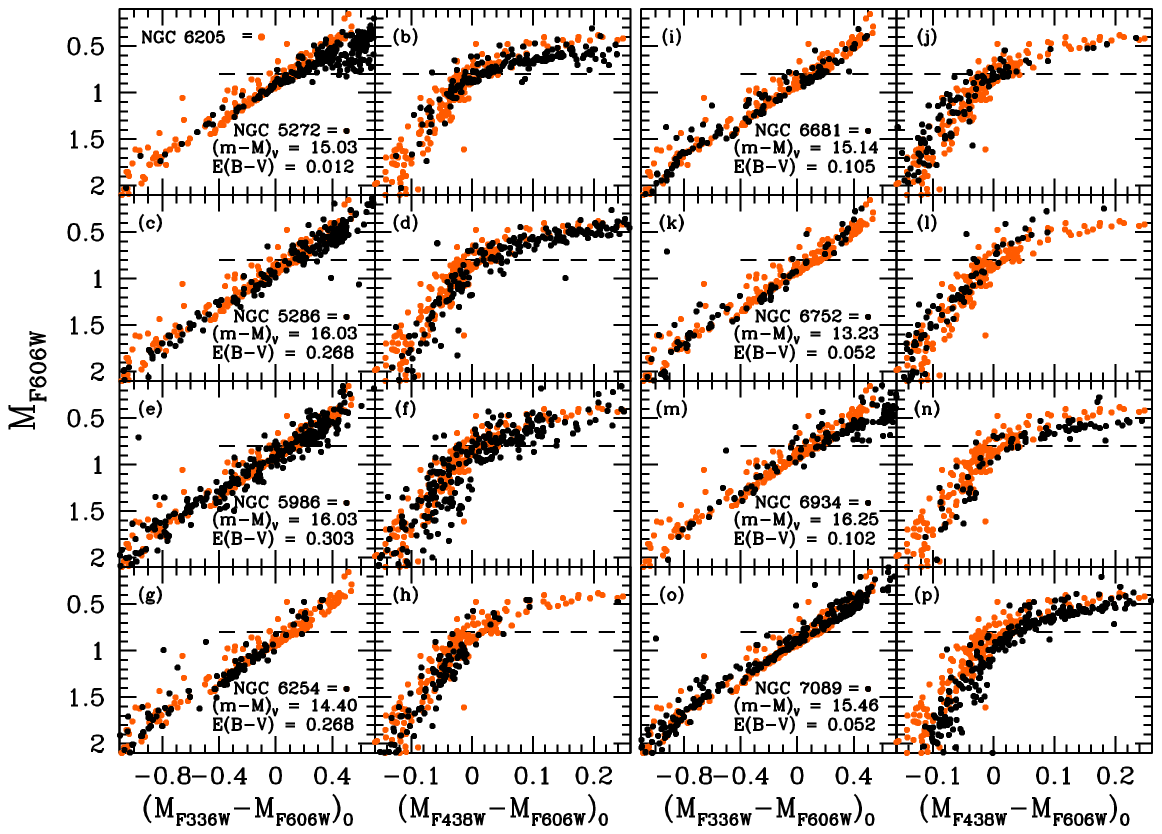}
\caption{As in Figs.~\ref{fig:f9} and~\ref{fig:f10}, except that the HBs of most
of the GCs with $-1.7 \lta$ [Fe/H] $\lta -1.5$ (see Fig.~\ref{fig:f11}) have
been superimposed on the HB of NGC$\,$6205 (M$\,$13), assuming $E(B-V) = 0.017$
and $(m-M)_V = 14.45$ for the latter and the specified cluster parameters for
the other GCs.}
\label{fig:f13}
\end{center}
\end{figure*}

The intercomparisons of blue HB populations on UV-optical CMDs provide valuable
consistency checks of the relative cluster properties that are implied by fits
of observed HB populations to ZAHB models.  In particular, such plots provide 
important connections between GCs that have, and those which do not have, red
HB components that are easily fitted to ZAHB models.  As shown in panel (a) of
Figure~\ref{fig:f13}, the blue HB stars in M$\,$13 and M$\,$3 superimpose each
other almost exactly if M$\,$13 has $E(B-V) = 0.017$ and $(m-M)_V = 14.45$ and
M$\,$3 has the indicated properties.  However, this does not rule out the
possibility that M$\,$13 has $E(B-V) = 0.024$ and $(m-M)_V = 14.42$ because an
increased reddening by 0.007 mag compensates for a 0.03 mag reduction in the
apparent distance modulus, thereby displacing the HB stars along the the same
line as the M$\,$3 HB.  To choose between these two or alternative
possibilities, it is necessary to know {\it a priori} the relative CMD locations
of the MS fiducials of M$\,$13 and M$\,$3 at a fixed absolute magnitude.  
 
The near coincidences of the the blue HB stars in NGC$\,$6681 and NGC$\,$6752
onto their counterparts in M$\,$13 (panels i, k) are really quite remarkable.
Even at $M_{F606W} > 1.5$, they lie on top of one another and exhibit exactly 
the same morphologies.   While there are no obvious difficulties in similar
plots for NGC$\,$6254 and NGC$\,$6934, one has the visual impression that better
matches of the NGC$\,$5286 and NGC$\,$5986 HBs to that of M$\,$13 would be
obtained if the former had higher reddenings and/or a larger distance moduli.
However, as already mentioned, such changes to the basic cluster parameters
would result in somewhat less satisfactory fits of the cluster HB stars to the
ZAHB models than those shown in Fig.~\ref{fig:f11}h,i.  Some of the scatter
of the NGC$\,$5286 HB can probably be attributed to the star-to-star metal
abundance variations that were recently detected by \citet{mmk15} or, in
the case of NGC$\,$5986, to the possible presence of differential reddening.
Regardless, the comparisons of the $(M_{F336W}-M_{F606W})_0$ colours of the
cluster HB stars that are presented in Fig.~\ref{fig:f13} are generally very
supportive of the adopted reddenings and distance moduli.  
%

\begin{figure*}
\begin{center}
\includegraphics[width=0.96\textwidth]{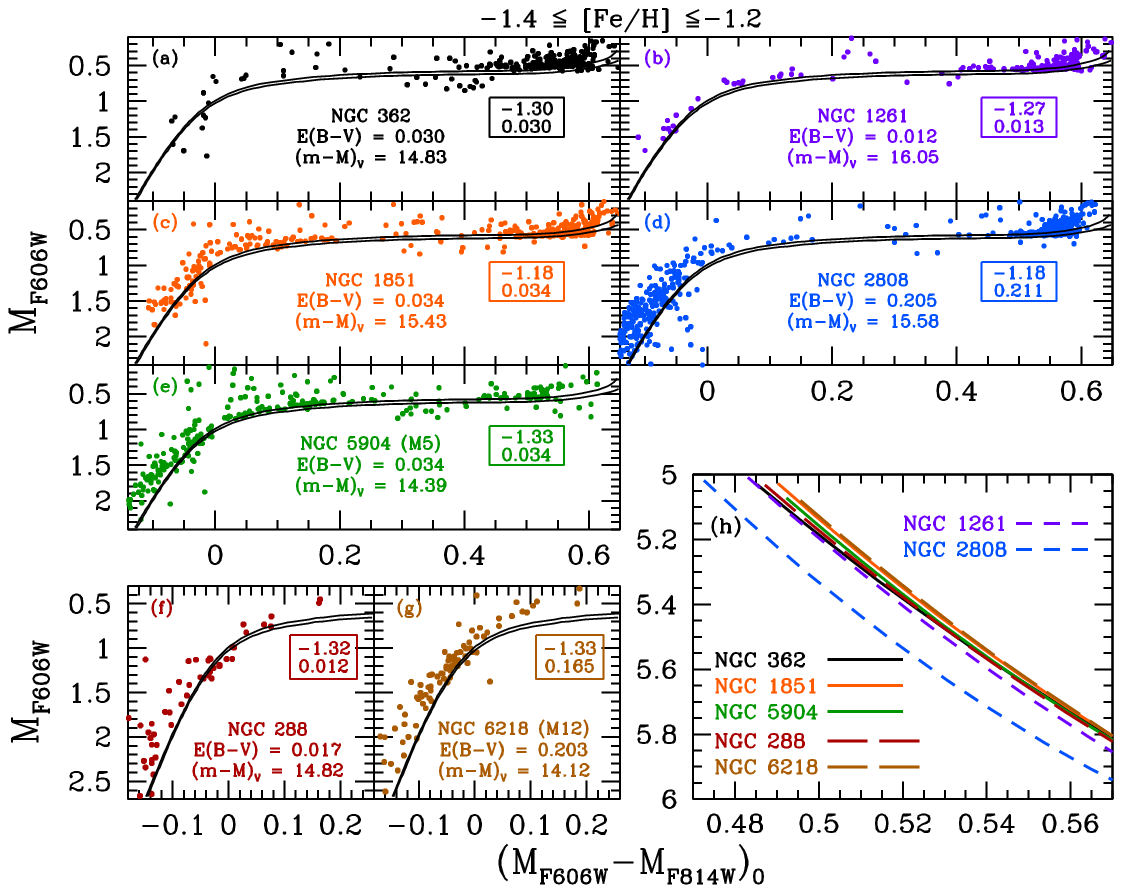}
\caption{As in Fig.~\ref{fig:f5} and~\ref{fig:f11}, except that the HB
populations of GCs with $-1.4 \lta$ [Fe/H] $\lta -1.2$ have been fitted to ZAHBs
for the minimum and maximum metallicities in these ranges.}
\label{fig:f14}
\end{center}
\end{figure*}

The same can be said of the $(M_{F336W}-M_{F438W})_0,\,M_{F606W}$ diagrams,
except for the one given in panel (p), which shows that there is a significant
offset between the M$\,$13 and M$\,$2 HBs.  This difficulty cannot be resolved
by simply increasing the assumed reddening of M$\,$2 because a higher value of
$E(B-V)$ would introduce a significant discrepancy between the cluster HBs in
panel (o).  Curiously, a similar comparison on the
$(M_{F606W}-M_{F814W})_0,\,M_{F606W}$ diagram (not shown) suggests that, if
anything, M$\,$2 has $E(B-V) < 0.052$; i.e., the discrepancy between the HB
stars in M$\,$13 and M$\,$2 is in the opposite sense, though only marginally.
Thus, whatever is responsible for the apparent separation of the HBs of M$\,$13
and M$\,$2 in panel (p) has something to do with the $F438W$ photometry.  The
apparent mismatch of the $(M_{F438W}-M_{F438W})_0$ colours might be due to
chemical abundance differences since M$\,$2 is an especially peculiar GC in
terms of the number of chemically distinct stellar populations that it contains
and the wide range of its chemical properties (\citealt{mmp15a}).  It should
also be kept in mind that small cluster-to-cluster differences in the
photometric zero points of the NLP18 photometry may have some impact on the
intercomparisons of UV-optical CMDs that are presented in this study. 

\subsection{GCs that have $-1.4 \lta$ [Fe/H] $\lta -1.2$}
\label{subsec:midhi}

The fitting of the HB populations in 5 GCs that have $-1.4 \lta$ [Fe/H] $\lta
-1.2$ to the relevant ZAHB sequences are presented in Figure~\ref{fig:f14}.
Aside from NGC$\,$288 and M$\,$12, they have significant numbers of red HB
stars that are easily matched to ZAHB models.  The inferred values of $(m-M)_V$
from such fits clearly have very little dependence on the assumed reddenings,
though the latter are needed for the determination of true distance moduli;
consequently, it is of some importance to adopt the best possible estimates of
the $E(B-V)$ values.  However, as for the more metal-deficient GCs that have
been considered in this study, quite agreeable results are obtained when the
reddenings from dust maps to within their $1\,\sigma$ uncertainties are adopted
for most of the clusters.  As shown in panel (h), the MS fiducial sequences of 
NGC$\,$362, NGC$\,$1261, NGC$\,$1851, and M$\,$5 define quite a narrow band
on the $(M_{F606W}-M_{F814W})_0,\,M_{F606W}$ diagram, which is to be expected
given they have nearly the same metallicities. 

\begin{figure*}
\begin{center}
\includegraphics[width=0.96\textwidth]{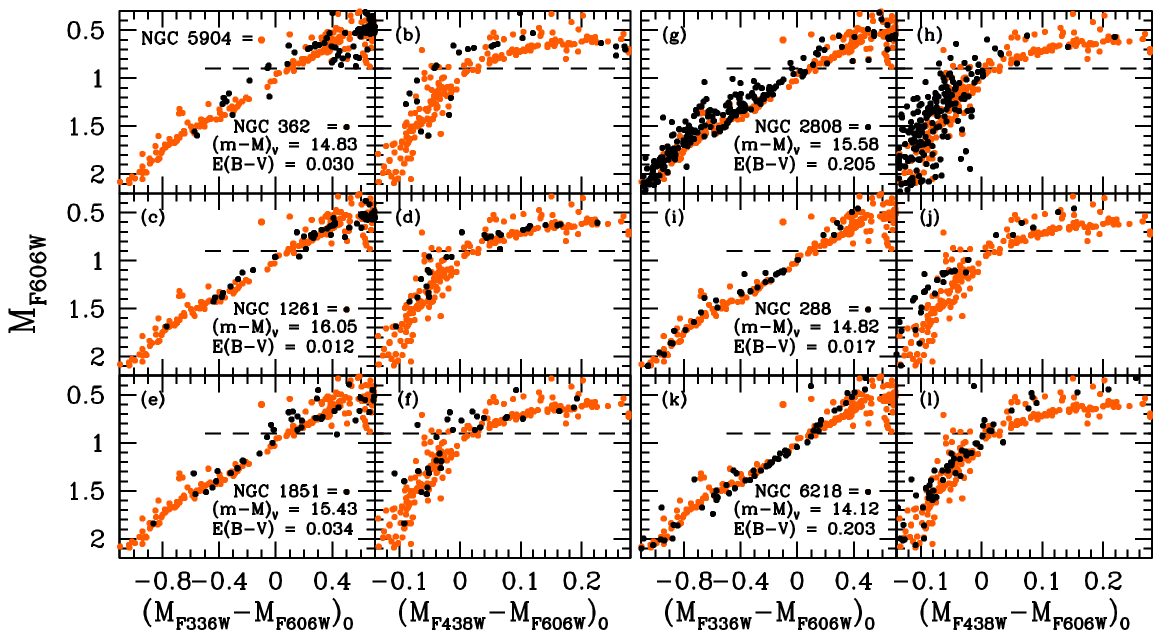}
\caption{As in Figs.~\ref{fig:f9},~\ref{fig:f10}, and~\ref{fig:f13}, except
that the HBs of GCs with $-1.4 \lta$ [Fe/H] $\lta -1.2$ (see Fig.~\ref{fig:f14})
have been superimposed on the HB of NGC$\,$5904 (M$\,$5), assuming $E(B-V) =
0.045$ and $(m-M)_V = 14.39$ for the latter and the indicated cluster parameters
for the other GCs.}
\label{fig:f15}
\end{center}
\end{figure*}

The MS of NGC$\,$2808 is the only one that is significantly offset from the
others, but this is to be expected as well.  NGC$\,$2808 is another Category 5
cluster with a very extended blue HB that has clearly had a very complex
chemical evolution history (see \citealt{mmp15b}).  Moreover, it has
spectroscopically confirmed He abundance variations by up to $\Delta\,Y = 0.09$,
relative to the primordial helium abundance (\citealt{mmp14}).  The MS of
NGC$\,$2808 should, therefore, be bluer than those of other GCs with similar or
even somewhat lower metallicities when the cluster HB is fitted to appropriate
ZAHB models, and this is precisely what is obtained when the dust-map reddening
(to within $1\,\sigma$) is adopted.  However, the offset of the blue dashed
curve relative to the others that have been plotted in Fig.~\ref{fig:f14}n is
larger than one would expect, particularly if NGC$\,$2808 has a lower 
metallicity than most of the other clusters by $\approx 0.1$ dex.  Better
consistency would be obtained if it has [Fe/H] $= -1.29$ (\citealt{ki03}) and a 
smaller reddening, say $E(B-V) = 0.200$, which is still within the $2\,\sigma$
error bar of the dust-map value.

The CMDs of NGC$\,$288 and M$\,$12 are much less straightforward to understand.
Because these Category 4 clusters have very blue HBs, their apparent distance
moduli as derived from fits of their HB stars to ZAHB models are very dependent
on the cluster reddenings.  If the $E(B-V)$ values from dust maps and
ZAHB-based distance moduli are adopted, the MS stars in NGC$\,$288 and M$\,$12
would be {\it redder} than their counterparts in M$\,$5, which has the same 
[Fe/H] value to within 0.01 dex according to CBG09.  This adds to the
considerable evidence presented in the previous section that second-parameter
clusters appear to have anomalously red main sequences for their metallicities.
In order for the MS fiducials of NGC$\,$288 amd M$\,$12 to be in close promixity
to those of the other GCs with [Fe/H] $\sim -2.3$, except NGC$\,$2808, which is
known to have high $Y$, it is necessary to adopt $E(B-V) \approx 0.017$ and
$(m-M)_V \approx 14.82$ for NGC$\,$288 and $E(B-V) \approx 0.203$ and $(m-M)_V
\approx 14.12$ for M$\,$12 (see panels f--h).  Larger reddenings by $\sim
0.01$ mag would be needed to produce the sufficiently blue main sequences that
would be expected if enhanced He abundances are primarily responsible for their
extremely blue HBs.

Valuable support for the adopted parameter values is provided by comparisons of
the cluster HBs on UV-optical CMDs.  Fortunately, all of the GCs in this group
have at least a few very blue HB stars; consequently, such intercomparisons may
be used to check the relative values of $E(B-V)$ and $(m-M)_V$.  In fact, the
fits of the reddest HB stars to ZAHB models in those GCs that have red HB
components should yield apparent distance moduli that are accurate in both an
absolute and relative sense.  As a result, the UV-optical CMDs primarily provide
helpful constraints on the relative reddenings of such clusters.  M$\,$5 is the
obvious choice for the reference cluster because its HB spans a very wide
colour range and because it has a relatively low reddening.  The various
$(M_{F336W}-M_{F606W})_0,\,M_{F606W}$ diagrams in Figure~\ref{fig:f15} show that
the bluest HB stars in NGC$\,$1261, NGC$\,$1851, and NGC$\,$2808 superimpose
their counterparts in M$\,$5 very well when the specified values of $E(B-V)$ and
$(m-M)_V$ are adopted and M$\,$5 has $E(B-V) = 0.034$ and $(m-M)_V =
14.39$.\footnote{\citet{gmk19} recently found from their analysis of photometry
in 29 passbands, ranging from the ultraviolet to the mid-infrared, that M$\,$5
has $E(B-V) = 0.054 \pm 0.02$.  Such a high reddening is very unlikely because
it would imply that the MS stars in M$\,$5 are {\it bluer}, at the same absolute
magnitude, than the MS stars in, e.g., M$\,$92 and M$\,$30, which have lower
[Fe/H] values by $\sim 1$ dex.  Consistency could be found if the distance
modulus of M$\,$5 were also increased by a large amount, but this would make its
HB much brighter than those of other clusters with similar metallicities.}  (As
noted above, NGC$\,$2808 may have a metallicity closer to [Fe/H] $= -1.3$ than
to $-1.2$, in which case, the fit of its HB to ZAHB models would yield $(m-M)_V
\approx 15.60$, and it would be necessary to adopt $E(B-V) \approx 0.200$ in
order to obtain very close to the same intercomparison of the NGC$\,$2808 and
M$\,$5 HBs that is shown in Fig.~\ref{fig:f15}g.) 

NGC$\,$362 has only a few blue HB stars that apparently straddle the M$\,$5 HB.
An improved fit of the four stars which lie just below the band that constitutes
the M$\,$5 HB could be obtained if the apparent distance modulus of NGC$\,$362
were increased by $\sim 0.02$ mag, though Fig.~\ref{fig:f14}a suggests that
$(m-M)_V = 14.85$ would be too high.  This slight inconsistency might
be due to small errors in the photometric zero points.  It is interesting that
most of the bluest HB stars in NGC$\,$362 are brighter than the majority of the
core He-burning stars in M$\,$5, presumably due to enhanced He abundances. 

\begin{figure}
\begin{center}
\includegraphics[width=0.96\columnwidth]{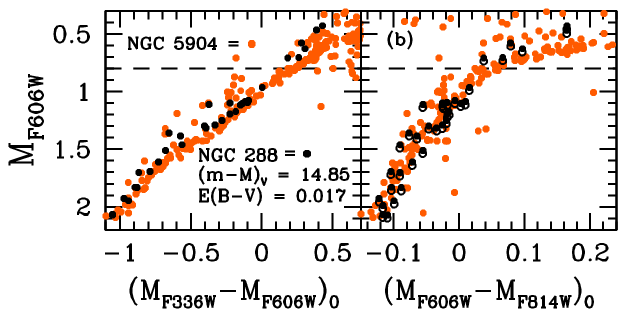}
\caption{Panel (a): as in panel (i) of the previous figure except that the
apparent distance modulus of NGC$\,$288 has been increased by 0.03 mag. Panel
(b): a comparison of the HBs of NGC$\,$288 and M$\,$5 on the
$(M_{F606W}-M_{F814W})_0,\,M_{F606W}$ diagram.  The small filled and open 
circles assume $E(B-V) = 0.017$ and $(m-M)_V = 14.85$ and 14.82, respectively,
for NGC$\,$288.}
\label{fig:f16}
\end{center}
\end{figure}

Panels (i) and (k) of Fig.~\ref{fig:f15} show that the blue HB populations of
NGC$\,$288 and NGC$\,$6218 (M$\,$12) superimpose the HB of M$\,$5 almost exactly
when the indicated cluster parameters are adopted.  The fact that there is very
little scatter about the tight, nearly linear distributions of their HB stars,
in stark contrast with NGC$\,$2808 (see panel g), suggests that they contain
chemically relatively simple stellar populations.  It is surprising that
NGC$\,$288, in particular, appears to require a higher reddening by $\approx
0.005$ mag than the dust-map determination, which has a $1\,\sigma$ uncertainty
of 0.001 mag, in order to obtain an acceptable fit of its HB stars to ZAHB
models simultaneously with a close match of its MS with those of M$\,$5 and
NGC$\,$362 (Fig.~\ref{fig:f14}f,h).  Although an even higher reddening and/or
an increased apparent distance modulus would tend to improve the fit of the HB
stars in NGC$\,$288 with $M_{F606W} \approx 1.2$ to ZAHB models, an increased
value of $(m-M)_V$ by as little as 0.03 mag results in a significant offset
between its HB and the HB of M$\,$5 --- as illustrated in Figure~\ref{fig:f16}a.
A  higher reddening by $\sim 0.007$ mag would have similar consequences, unless
the apparent distance modulus is also decreased by about 0.03 mag.  The
parameter values that were assumed in Fig.~\ref{fig:f15}i are clearly preferable.

\begin{figure*}
\begin{center}
\includegraphics[width=0.96\textwidth]{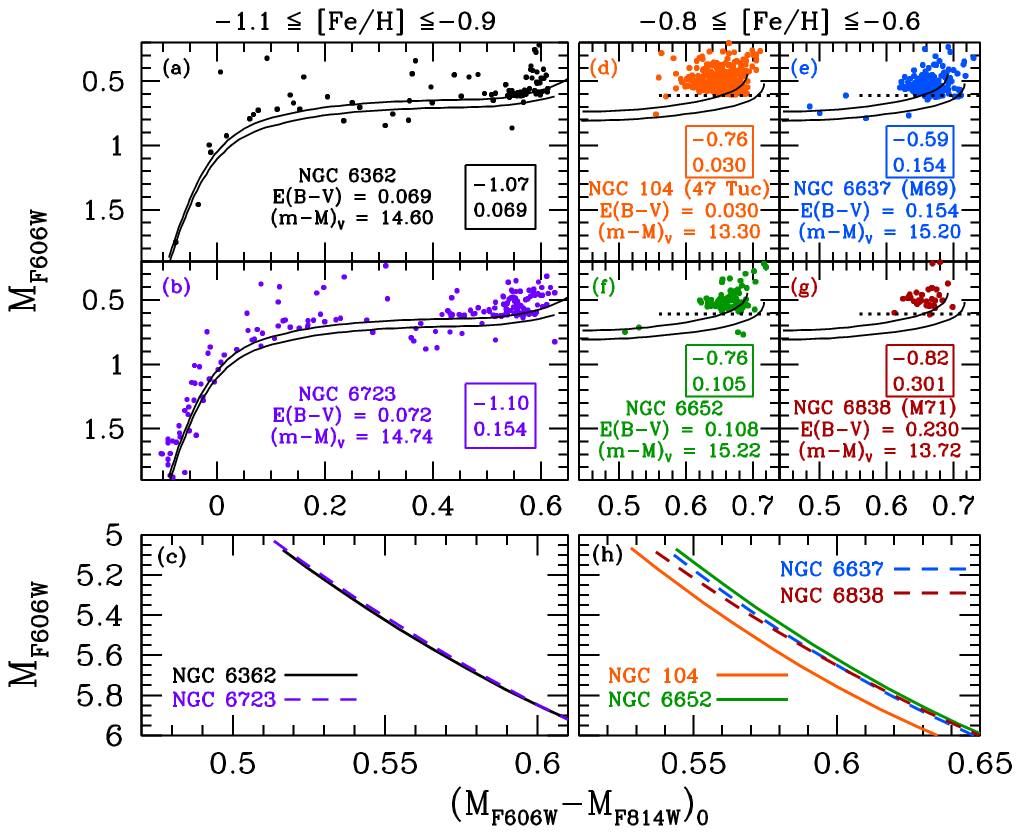}
\caption{As in Figs.~\ref{fig:f5},~\ref{fig:f11}, and~\ref{fig:f14}, except that
the HB populations of GCs with $-1.1 \lta$ [Fe/H] $\lta -0.9$ (panels a and b)
and $-0.8 \lta$ [Fe/H] $\lta -0.6$ (panels d--g) have been fitted to ZAHBs for
the minimum and maximum metallicities in these ranges.  The
faintest HB stars in 47 Tuc are represented by the horizontal dotted line (see
panel d).  This line is reproduced in panels (c)-(g) to facilitate comparisons
with the faintest stars in M$\,$69, NGC$\,$6652, and M$\,$71.}
\label{fig:f17}
\end{center}
\end{figure*}


For the most part, the comparisons of cluster HBs on
$(M_{F438W}-M_{F606W})_0,\,M_{F606W}$ diagrams support the adopted cluster
parameters, though visual inspections of panels (d) and (j) suggest that 
the $E(B-V)$ values of NGC$\,$1261 and NGC$\,$288 may be too high.  However,
as in the case of M$\,$2, which showed a similar discrepancy but in the opposite
sense relative to the reference cluster M$\,$13 (see Fig.~\ref{fig:f13}l), no
such problems are found when the cluster HBs are compared on
$(M_{F606W}-M_{F814W})_0,\,M_{F606W}$ diagrams.  An example of such a plot is 
presented in panel (b) of Fig.~\ref{fig:f16}, which shows that the bluest HB
stars in M$\,$5 and NGC$\,$288 superimpose each other exceedingly well,
regardless of small differences in the assumed distance moduli.  Evolutionary
effects are almost certainly responsible for the enhanced brightnesses of the
reddest HB stars in NGC$\,$288 relative to the stars in M$\,$5 with the same
intrinsic colours.  

The discrepant $(M_{F438W}-M_{F606W})_0$ colours are apparently connected with
the $F438W$ photometry.  While this problem would be well worth investigating,
it does not call into question the results of this investigation, which rely
primarily on $F336W,\,F606W,\,F814W$ observations.  

\subsection{GCs that have $-1.1 \lta$ [Fe/H] $\lta -0.6$}
\label{subsec:hiz}

The fits to ZAHB models of the HB populations of the most metal-rich GCs that
have been considered in this investigation are shown in Figure~\ref{fig:f17}.
Since most of the stars are concentrated in red clumps, it is a relatively
straightforward to derive their apparent distance moduli on the assumption of
well-supported estimates of their metallicities, though the resultant values
of $(m-M)_V$ for the clusters with the reddest HBs will have larger 
uncertainties because their HB stars are matched to the portions of ZAHBs that
bend upwards at their red ends.  As a result, such fittings have an increased
dependence on the uncertainties associated with the reddenings and with the
colours predicted by the stellar models.

NGC$\,$6362 and NGC$\,$6723 have almost the same [Fe/H] values according to
CBG09 and if the dust-map reddening is adopted for NGC$\,$6362 and $E(B-V) =
0.072$ for NGC$\,$6723, their MS fiducials are nearly coincident if the
ZAHB-based distance moduli are also adopted, as shown in panel (c).  VBLC13
found from a similar fitting of the NGC$\,$6723 HB to their ZAHB models that the
reddening of this GC according to dust maps has to be too large by about a
factor of two; also see \citet{gkm23}, who derived $E(B-V) = 0.068 \pm 0.01 \pm
0.02$ (statistical and systematic uncertainties) for NGC$\,$6723 from an
analysis of UV to mid-infrared observations.

With regard to more metal-rich systems, there are other concerns aside from
those mentioned above.  Because the effects of $\pm 0.1$ dex uncertainties in
GC metallicities on the predicted locations of upper-MS stars increase with
increasing [Fe/H], the superposition of the fiducial sequences for the cluster
upper-MS stars provides a progressively weaker constraint on the reddenings and
apparent distance moduli as the metallicity increases.  Moreover, comparisons
of the cluster HBs on $(M_{F336W}-M_{F606W})_0,\,M_{F606W}$ diagrams cannot be
used to constrain the basic properties of clusters with with [Fe/H] $\gta -1.0$
because such systems contain very few, if any, HB stars on the blue side of the
instabiity strip.  Fortunately, 47 Tucanae is sufficiently well understood that
it can be used as a bridge between clusters of both lower and higher
metallicities.

The analysis of the eclipsing binaries in 47 Tuc by \citet{bvb17} indicated
a preference for [Fe/H] $\approx -0.70$ instead of the CBG09 determination
of $-0.76$. In fact, the higher metallicity is quite possibly
the best available estimate given that \citet{ki03} obtained [Fe/H] $= -0.70$
from the equivalent widths of Fe II lines, which are much less affected, if at
all, by departures from local thermodynamic equilibrium than Fe I lines.
Brogaard et al.~also concluded from their study of the binary V69 that 47 Tuc
has $(m-M)_V = 13.30$.  This is only 0.03 mag smaller than the distance modulus
that was subsequently derived from {\it Gaia} DR2 parallaxes by \citet{crc18}
if $E(B-V) = 0.030$ (from dust maps) is adopted.  On the other hand, a somewhat
lower value, $(m-M)_V = 13.27$, is favoured by simulations of the cluster HB
that successfully reproduce the observed HB on the
$(M_{F606W}-M_{F814W})_0,\,M_{F606W}$ diagram if the He abundance is assumed to
vary from $Y = 0.257$ to 0.287 (see DVKF17).  As shown in Fig.~\ref{fig:f17}d,
$(m-M)_V = 13.30$ and $E(B-V) = 0.030$ results in a good fit of the faintest
and reddest HB stars in 47 Tuc to ZAHB models for [Fe/H] $\approx -0.70$. 

The horizontal dotted line that has been drawn through these stars in panel (d)
has been reproduced in panels (e)--(g).  Since M$\,$69 is only slightly more
metal rich than 47 Tuc, those HB stars that are closest to their ZAHB locations
should lie just below the dotted line, resulting in $(m-M)_V \approx 15.20$.  It
turns out that, if $E(B-V) = 0.154$ (from dust maps) is also assumed, the
reddest of these stars have colours that are similar to the ZAHB models for
[Fe/H] $= -0.60$, which is consistent with the metallicity of M$\,$69 that was
derived by CBG09.  Similarly, the faintest HB stars in NGC$\,$6652 and in
M$\,$71 should be matched to the dotted line or located slightly above it,
respectively, given that these clusters appear to have the same or a somewhat
lower [Fe/H] value than 47 Tuc.  Although the assumption of the dust-map
reddening to within its $1\,\sigma$ uncertainty leads to a satisfactory match
of the HB stars in NGC$\,$6652 to ZAHB models (panel f), the reddening of
M$\,$71 has to be much less than the dust-map determination in order to obtain
a similar result for this cluster (panel g).  Indeed, a reddening near $E(B-V)
= 0.23$ is required in order for its MS stars to have similar CMD locations
as those of other GCs in this group, as shown in panel (h).  A higher median
He abundance due to chemical self-enrichment presumably explains most of the
blueward extension of the MS of 47 Tuc relative to the others.

\section{On the Importance of Stellar Rotation and Mass Loss in GCs}
\label{sec:second}

Some further discussion is warranted concerning the finding that, when the best
estimates of the reddenings from dust maps are adopted, the GCs with the bluest
HBs in many of the second-parameter pairs, including M$\,$3--M$\,$13 and
NGC$\,$288--NGC$\,$362, have somewhat {\it redder} main sequences than the
clusters in each pair with HBs that are more typical of their shared
metallicities.  Cluster-to-cluster differences in the distributions of stellar
rotation can potentially explain this anomaly (see \citealt{mg76}), though more
recent stellar models that incorporate a much more sophisticated treatment of
rotation are not supportive of this possibility (see \citealt{ddp89}).  While
there is little doubt of the importance of rotation in modulating the mass loss
that occurs during RGB evolution (\citealt{ren77}) and/or at the helium flash
(\citealt{pet82}), thereby impacting HB morphologies, it is not expected to have
any effect on the temperatures of cluster MS stars.  On the other hand, the
modeling of rotation involves free parameters that need to be calibrated, and
various assumptions must be made concerning the transport of angular momentum
in stellar interiors, the loss of angular momentum via stellar winds,
etc.~(see, e.g., \citealt{pdd91}).  Consequently, one cannot say with absolute
certainty that the median fiducial sequences for the MS stars in all GCs,
especially some of those with extended blue HBs, are unaffected by rotation. 

To be sure, it is an important prediction of the models computed by 
\citet{pdd91} that metal-poor stars are able to retain sufficient internal
angular momentum to explain the rapid rotations that have been determined in
several of the blue HB stars in M$\,$13 (\citealt{pet83}) and in NGC$\,$288
(\citealt{pet85}).  M$\,$13 is quite a remarkable cluster in this regard insofar
as approximately one-third of the 32 HB stars from the \citet{pet83} and
\citet{prc95} surveys, with temperatures in the range $7000 < \teff \lta 11,000$
K, have $20 \lta \vsini \lta 40$ km$\,$s$^{-1}$; the mean value of $\vsini$ for
the entire sample is 18.1 km$\,$s$^{-1}$.  By comparison, \citet{pet85} obtained
values of $\vsini$ ranging from 12 to 22 km$\,$s$^{-1}$ ($\langle \vsini\rangle =
16.3$ km$\,$s$^{-1}$) for six of the seven HB stars that she studied, all of
which have temperatures between 7000 and 9000 K.   However, all 16 stars with
$9000 < \teff < 11,000$~K that were subsequently investigated by \citet{prc95}
have $\vsini \lta 12$ km$\,$s$^{-1}$, resulting in $\langle \vsini\rangle =
7.5$ km$\,$s$^{-1}$. 

It is not clear what to make of the apparent variation of $\langle \vsini\rangle$
with $\teff$ in NGC$\,$288, as there is no indication of a similar trend along 
the M$\,$13 HB.  Regardless, the rotational velocities in NGC$\,$288 are
much smaller than in M$\,$13 and, in fact, they are not very different from the
rotation rates that have been determined for the 22 blue HB stars in M$\,$3
that were studied by \citet{prc95}; they have temperatures that are mostly
between 7000 and 9000~K (like the cooler group of stars in NGC$\,$288) and 
$\vsini$ values that range from 2 to 21 km$\,$s$^{-1}$ ($\langle \vsini\rangle =
12.4$ km$\,$s$^{-1}$).  This would seem to argue against
rotation being the main driver of the very blue HB of NGC$\,$288.  However, the
latter has a significantly higher metallicity than M$\,$13 and M$\,$3, and more
metal-rich giants can be expected to lose more mass than those of lower [Fe/H],
perhaps especially if they are rotating, because they evolve along cooler tracks
and they reach higher luminosities prior to the He flash.  According to, e.g.,
the empirically based \citet{rei75} formula, which is still widely used in
stellar evolutionary computations (e.g., \citealt{phc21}), the rate at which
low-mass giants lose mass is proportional to $L^{1.5}/\teff^2$.  Indeed, if
rotation is of sufficient importance, it could delay the He flash
(\citealt{mg76}), which would allow more mass to be lost (\citealt{ren77}) and
possibly result in somewhat brighter HBs as the result of increased helium core
masses at the RGB tip (\citealt{sc98}).

As shown by \citet{rpa02} and particularly by \citet{bb03}, globular cluster
HB stars with $\teff \gta 11,000$--11,500~K are slow rotators with few, if any,
exceptions.  The fairly sharp transition at this temperature from relatively 
high values of $\vsini$ in cooler HB stars to $\lta 5$ km$\,$s$^{-1}$ in hotter
stars --- see, e.g., Behr's results for M$\,$13 and NGC$\,$288 in his Fig.~23
--- occurs at the same $\teff$ as the so-called ``Grundalh jump", where
radiative levitations greatly enhance the surface abundances of the metals,
causing a jump in the Str\"omgren $u$ magnitude (see \citealt{gcl99}).  This
feature is common to all GCs with HB stars that have $\teff \gta 11,500$~K,
irrespective of metallicity.  As suggested by \citet{vs02}, the increased metal
abundances can be expected to result in enhancements of both the mass-loss rates
and the concomitant losses of angular momentum via stellar winds.  Thus, the
hottest HB stars along blue tails that are now observed to be slow rotators
probably had much higher rotational velocities at the beginning of their core
He-burning lifetimes.  

On the other hand, some GCs with extended blue HB populations do not seem to
contain any fast rotators with $\teff < 11,500$~K, including NGC$\,$2808 and
NGC$\,$6093 (M$\,$80); see \citet{rpa02}.  In the case of NGC$\,$2808, the
spectroscopic study by \citet{mmp14} found that helium varies by up to
$\Delta\,Y \approx 0.09$, which can explain the extended HB of this Category 5
cluster.   As there are no indications from chromosome maps that such large He
abundance variations exist in M$\,$80 (\citealt{mmr18}), which is also a
Category 5 GC, it is possible that both rotation and helium (to a lesser extent
than in NGC$\,$2808) play a role in producing its HB.  Recio-Blanco et
al.~included only three HB stars with $\teff \lta 10,000$~K in their survey, and
since it is not uncommon to find a wide range in the measured $\vsini$ values
in some GCs (see \citealt{bb03}), it would be worthwhile to measure the rotational
velocities of many more HB stars in this cluster, especially those with lower
temperatures.


It is also noteworthy that rotation has been found to be more important in
M$\,$92 than in M$\,$68 or in M$\,$15 (\citealt{bb03}), which has the most
extended HB of the three GCs.  However, just as NGC$\,$2808 has a much
bluer MS than other clusters of similar metallicity with redder HB populations,
when the dust-map reddening to within 1--$2\,\sigma$ is
adopted, so is the MS of M$\,$15 significantly bluer than those of M$\,$92 and
M$\,$68 if dust-map determinations of $E(B-V)$ are adopted for all three
clusters (see Fig.~\ref{fig:f5}g). This is consistent with M$\,$15 having
significantly higher $Y$ than the other two GCs, as found from analyses of
chromosome maps (\citealt{mmr18}) --- which also suggest that M$\,$92 has a
somewhat higher He abundance, in the mean, than M$\,$68.   Although this could
explain why the HB of M$\,$92 has a greater blueward extension than that of
M$\,$68, the observed differences in their rotational properties and the
associated mass loss could also be a factor.  Indeed, if M$\,$92 has a higher
He abundance, why does it have a slightly redder MS than M$\,$68 when dust-map
reddenings are adopted for both GCs?  Although the answer to this question could
simply be that dust maps underestimate the reddening of M$\,$92, it would be
surprising that the dust-map $E(B-V)$ value is problematic for the cluster in
the most metal-deficient group that seems to have the highest fraction of
rapidly rotating blue HB stars, but not for any other GC in that group (see
Fig.~\ref{fig:f5}).
 

The cluster that is closest to being a ``smoking gun" with regard to the notion
that rotation and mass loss may be primarily responsible for the very blue HBs
in several clusters is NGC$\,$288.  Because this system has a higher
metallicity than most of the GCs with extremely blue HBs, its RGB stars must
undergo especially large amounts of mass loss in order to wind up on the blue
HB after the helium flash (recall Fig.~\ref{fig:f2}).  At $M_{F606W} \approx
1.0$, which is near the top of the blue HB of NGC$\,$288 (see
Fig.~\ref{fig:f14}f), a ZAHB star is predicted to have a mass of $0.597 \msol$,
according to the present models for [Fe/H] $= -1.3$, $Y = 0.252$, [O/Fe] $=
+0.6$, and [$m$/Fe] $= +0.4$ for all other $\alpha$ elements.  This may be
compared with a mass of $0.876 \msol$ or $0.812 \msol$, which are the predicted
masses of RGB tip stars at an age of 10.0 Gyr or 13.0 Gyr, respectively, if no
mass loss occurs.  (The difference in mass due to an age difference as large as
3.0 Gyr is clearly largely irrelevant.)  NGC$\,$288 has few HB stars fainter than
$M_{F606W} \approx 2.7$, where the predicted ZAHB mass is $0.542 \msol$.  Hence,
the entire observed HB of NGC$\,$288 can be explained if stars in this cluster
lose between $\approx 0.25$ and $\approx 0.31 \msol$ prior to reaching the HB,
assuming a mass of $0.85 \msol$ for the RGB tip precursors.  

The very tight sequence of the blue HB stars in NGC$\,$288 on the UV-optical
CMD (Fig.~\ref{fig:f15}i) provides compelling evidence that this cluster
contains chemically relatively simple stellar populations.  Its CMD is in stark
contrast with that of NGC$\,$2808 (Fig.~\ref{fig:f15}g), which has large 
star-to-star He abundance variations and other chemical peculiarities
(\citealt{mmp14}).  \citet{lsb18} also found from their examination of
chromosome maps that stars in NGC$\,$288 appear to have very homogeneous helium
and light element abundances.  Since age can play no more than a minor role in
producing blue HBs at higher metallicities (as noted above), mass loss,
presumably facilitated by high rotational velocities, would appear to be the
primary cause of the observed HB of NGC$\,$288.  (The same can be said of
M$\,$12, given the close similarity of its HB with that of NGC$\,$288; see
Fig.~\ref{fig:f15}k.)

DVKF17 have already shown that at least part of the explanation for the very
different HB morphologies of M$\,$3 and M$\,$13 is a large difference in the
mass that is lost from the giants in the two clusters.  According to the present
models for [Fe/H]  $\sim -1.55$, which is approximately the metallicity of
M$\,$13, its RGB stars would need to lose between $\sim 0.20$ and $\sim 0.30
\msol$ in order to account for the observed HB.  Although DVKF17 also showed
that the full extension of the M$\,$13 HB can be explained if the helium
abundances of member stars vary by  $\Delta\,Y \sim 0.08$, as compared with
$\Delta\,Y \lta 0.01$ in the case of M$\,$3, this possibility is not supported
by analyses of chromosome maps (\citealt{mmr18}) or by the fact that the MS
stars in M$\,$13 are nearly coincident with, if not slightly redder than, the
MS stars in M$\,$3, at the same absolute magnitudes, if it has $E(B-V) = 0.017$
(from dust maps).  The same difficulty applies to NGC$\,$6681 and NGC$\,$6752,
which have M$\,$13-like HBs (see Fig.~\ref{fig:f11}).  Thus, insofar as the CMDs
of at least these three GCs, as well as NGC$\,$288, are concerned, mass loss
would appear to be the most important ``second parameter".


However, there are some observations that appear to favour a much higher
reddening for M$\,$13 than the dust-map estimate.  As reported by DVKF17,
stellar models are unable to predict the periods of the mostly $c$-type RR Lyrae
variables in M$\,$13 satisfactorily, even if they have $E(B-V) = 0.025$ and
$(m-M)_V = 14.42$, which are close to the parameter values that are needed to
reconcile the CMDs of M$\,$3 and M$\,$13 with the differences in $Y$ that are
implied by HB simulations.  (No such difficulties were found in the analyses of
the M$\,$3 RR Lyrae stars in the previous study by \citealt{vdc16}.)  Because
pulsation periods vary inversely with $\teff$ and directly with luminosity
(\citealt{mcb15}), improved consistency would be obtained if a higher
reddening and/or a smaller distance modulus were assumed for M$\,$13.  Clearly,
the smaller value of $E(B-V) = 0.017$ from dust maps and the consequent
ZAHB-based distance modulus that has been derived in this investigation,
$(m-M)_V = 14.45$, exacerbate the problem.

\begin{figure}
\begin{center}
\includegraphics[width=\columnwidth]{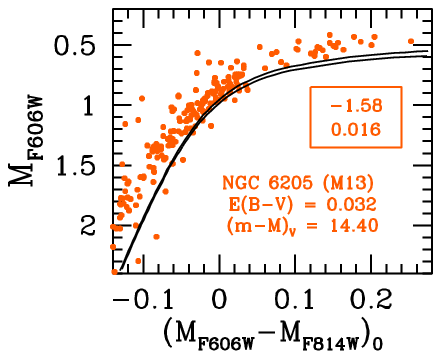}
\caption{Similar to Fig.~\ref{fig:f11}j except that a higher reddening and a
smaller value of $(m-M)_V$ have been assumed, as indicated.}
\label{fig:f18}
\end{center}
\end{figure}

Fortunately, the permitted values of the basic cluster parameters are
constrained by the intercomparisons of their respective HB populations on
UV-optical CMDs.  For instance, it would generally not be possible to obtain
a near coincidence of the blue HB stars in M$\,$3 and M$\,$13, similar to that
shown in Fig.~\ref{fig:f13}a on the assumption of smaller values of $(m-M)_V$
for M$\,$13, unless suitable increases in the adopted reddening are also
assumed.  In fact, a visually indistinguishable overlay of the M$\,$3 and
M$\,$13 HBs would be obtained if M$\,$13 has $E(B-V) = 0.032$ and $(m-M)_V =
14.40$, which would also provide a reasonably satisfactory resolution of the RR
Lyrae problem.  Of course, these results assume that the properties of M$\,$3
are well determined, though similar conclusions are obtained if other GCs are
considered.  In particular, the apparent distance modulus of M$\,$2 cannot be
much smaller than $(m-M)_V = 15.46$ without causing an unsatisfactory fit of
its reddest HB stars to the ZAHB models (see Fig.~\ref{fig:f11}m).  Reduced
distance moduli for NGC$\,$6681 and NGC$\,$6752 are also unlikely because the
adopted values of $(m-M)_V$ already imply that they are among the oldest of 
the Galactic GCs (see the next section).

Although the high reddening and short distance modulus seem to be preferred 
from the perspective of the M$\,$13 RR Lyrae variables, they result in a less
than satisfactory fit of the cluster HB stars to ZAHB models at $M_{F606W} >
1.3$, as illustrated in Figure~\ref{fig:f18} and highlighted by a comparison of
this plot with Fig.~\ref{fig:f11}j.   In addition, they cause a blueward offset
of the MS of M$\,$13 relative to that of M$\,$3 by $\sim 0.02$ mag, making it
considerably bluer than the M$\,$2 MS despite the latter having a lower [Fe/H]
value and CMD locations of its MS stars that are
consistent with enhanced He abundances (see Fig.~\ref{fig:f11}n).  Such a large
offset would be comparable with the separation in colour between the MS portions
of isochrones that have been computed for $Y = 0.25$ and 0.32, which is larger
by about a factor of two than the difference in the mean He abundances of M$\,$3
and M$\,$13 according to the HB simulations carried out by DVKF17.  This argues
against the parameter values that were assumed in the contruction of
Fig.~\ref{fig:f18}.  It is more likely that the actual reddening and
distance modulus of M$\,$13 lie within the ranges $0.016 \le E(B-V) \le 0.024$
and $14.42 \le (m-M)_V \le 14.45$, as favoured by DVKF17 and the present study.

Given that stellar models are generally quite successful in reproducing the
pulsational and evolutionary properties of globular cluster RR Lyrae stars (see,
e.g., \citealt[2018]{vdc16}), why do they fail to do so in the case of M$\,$13?
Is this is another consequence of the high number of rapid rotators in this
system?  Indeed, it would be worthwhile to explore the implications of rotating
HB models, not only for variable stars, but also for fits of GC HB
populations to ZAHB models.  The matching of the NGC$\,$288 HB to the ZAHBs, in
particular, is especially problematic because a relatively large number of the
HB stars with $M_V \approx 1.25$ lie below the model computations (see
Fig.~\ref{fig:f14}f).  It is quite possible that the adopted reddening is too
low or that there is a problem with the photometry, but stellar rotation might
also have some impact on the luminosities, temperatures, and colours of these
stars.

\section{Summary and Discussion}
\label{sec:sum}

New grids of ZAHB models for chemical abundances that are relevant to the
Galactic GCs have been presented in this study.  They have been applied to
the HB populations of 37 clusters in order to derive their apparent distance
moduli.  The very weak dependence of the location of MS stars on the
$(M_{F606W}-M_{F814W})_0,\,M_{F606W}$ diagram, especially at the lowest
metallicities, have been used to constrain the absolute values of $(m-M)_V$.
Comparisons of the bluest HB stars on UV-optical CMDs have been found to
provide particularly valuable constraints on the {\it relative} apparent
distance moduli of clusters of similar [Fe/H].   Although it was not anticipated
at the outset of this investigation, good fits of the observed HBs to ZAHB
sequences could be obtained for $\sim 75$\% of the GCs on the assumption of
dust-map reddenings to within their $1\,\sigma$ uncertainties.  This provides
quite a strong indication that the $E(B-V)$ values from dust maps are generally
very accurate.

The main results of this study are listed in Table~\ref{tab:t3}.  This gives,
for the GCs that are identified in the first two columns, the best estimates of
the reddenings and their uncertainties from dust maps, the adopted values of
$E(B-V)$, and the derived ZAHB-based apparent distance moduli.  True distance
moduli can be readily calculated from $(m-M)_0 = (m-M)_V - R_V\,E(B-V)$, with
$R_V = 3.13$ (\citealt{cv14}) using the reddenings given in the fourth
column.  They differ significantly from dust-map determinations only in the
case of a few moderately reddened systems, such as NGC$\,$3201 and NGC$\,$6723,
for which the adopted values of $E(B-V)$ are expected to be more accurate.  The
resultant values of $(m-M)_0$ have not been tabulated because they are less
reliable than the apparent distance moduli that have been determined in this
study as a result of their greater dependence on $E(B-V)$.  For instance, the
$(m-M)_V$ values that are found from the fitting to ZAHB models of the HB
populations in GCs that have significant numbers of red HB stars depend only
weakly, if at all, on the reddening. 

\begin{table}
\centering
\caption{Summary of the Results} 
\label{tab:t3}
\smallskip
\begin{tabular}{ccccc}
\hline
\hline
\noalign{\smallskip}
  NGC & Other & $E(B-V)$ & $E(B-V)$ & $(m-M)_V$ \\
  Number & Name & (dust maps) & (adopted) & (derived) \\
\noalign{\vskip 1pt}
\hline
\noalign{\smallskip}
 ~104 & 47 Tuc  & $0.030 \pm 0.001$ & 0.030 & 13.30 \\ 
 ~288 &         & $0.012 \pm 0.001$ & 0.017 & 14.82 \\ 
 ~362 &         & $0.030 \pm 0.001$ & 0.030 & 14.83 \\ 
 1261 &         & $0.013 \pm 0.001$ & 0.012 & 16.05 \\ 
 1851 &         & $0.034 \pm 0.002$ & 0.034 & 15.43 \\ 
 2808 &         & $0.211 \pm 0.006$ & 0.205 & 15.58 \\ 
 3201 &         & $0.237 \pm 0.016$ & 0.273 & 14.17 \\ 
 4147 &         & $0.024 \pm 0.001$ & 0.024 & 16.45 \\ 
 4590 & M$\,$68 & $0.057 \pm 0.001$ & 0.057 & 15.24 \\ 
 5024 & M$\,$53 & $0.019 \pm 0.001$ & 0.019 & 16.38 \\ 
 5053 &         & $0.016 \pm 0.001$ & 0.016 & 16.22 \\ 
 5272 & M$\,$3  & $0.012 \pm 0.001$ & 0.012 & 15.03 \\ 
 5286 &         & $0.282 \pm 0.014$ & 0.268 & 16.03 \\ 
 5466 &         & $0.016 \pm 0.001$ & 0.017 & 16.05 \\ 
 5904 & M$\,$5  & $0.034 \pm 0.001$ & 0.034 & 14.39 \\ 
 5986 &         & $0.313 \pm 0.010$ & 0.303 & 16.03 \\ 
 6101 &         & $0.097 \pm 0.003$ & 0.115 & 16.11 \\ 
 6205 & M$\,$13 & $0.016 \pm 0.001$ & 0.017 & 14.45 \\ 
 6218 & M$\,$12 & $0.165 \pm 0.002$ & 0.203 & 14.12 \\ 
 6254 & M$\,$10 & $0.268 \pm 0.007$ & 0.268 & 14.40 \\ 
 6341 & M$\,$92 & $0.021 \pm 0.001$ & 0.022 & 14.72 \\ 
 6362 &         & $0.069 \pm 0.001$ & 0.069 & 14.60 \\ 
 6397 &         & $0.175 \pm 0.003$ & 0.175 & 12.49 \\ 
 6541 &         & $0.145 \pm 0.011$ & 0.135 & 14.75 \\ 
 6584 &         & $0.099 \pm 0.006$ & 0.091 & 15.92 \\ 
 6637 & M$\,$69 & $0.154 \pm 0.002$ & 0.154 & 15.20 \\ 
 6652 &         & $0.105 \pm 0.003$ & 0.108 & 15.22 \\ 
 6681 & M$\,$70 & $0.100 \pm 0.005$ & 0.105 & 15.14 \\ 
 6723 &         & $0.154 \pm 0.027$ & 0.072 & 14.74 \\ 
 6752 &         & $0.052 \pm 0.001$ & 0.052 & 13.23 \\ 
 6809 & M$\,$55 & $0.127 \pm 0.001$ & 0.120 & 13.95 \\ 
 6838 & M$\,$71 & $0.301 \pm 0.016$ & 0.230 & 13.72 \\ 
 6934 &         & $0.098 \pm 0.003$ & 0.102 & 16.25 \\ 
 6981 & M$\,$72 & $0.054 \pm 0.002$ & 0.054 & 16.25 \\ 
 7078 & M$\,$15 & $0.101 \pm 0.003$ & 0.101 & 15.41 \\ 
 7089 & M$\,$2  & $0.042 \pm 0.001$ & 0.052 & 15.46 \\ 
 7099 & M$\,$30 & $0.047 \pm 0.002$ & 0.047 & 14.75 \\ 
\noalign{\smallskip}
\hline
\noalign{\smallskip}
\end{tabular}
\end{table}

\begin{figure}
\begin{center}
\includegraphics[width=0.96\columnwidth]{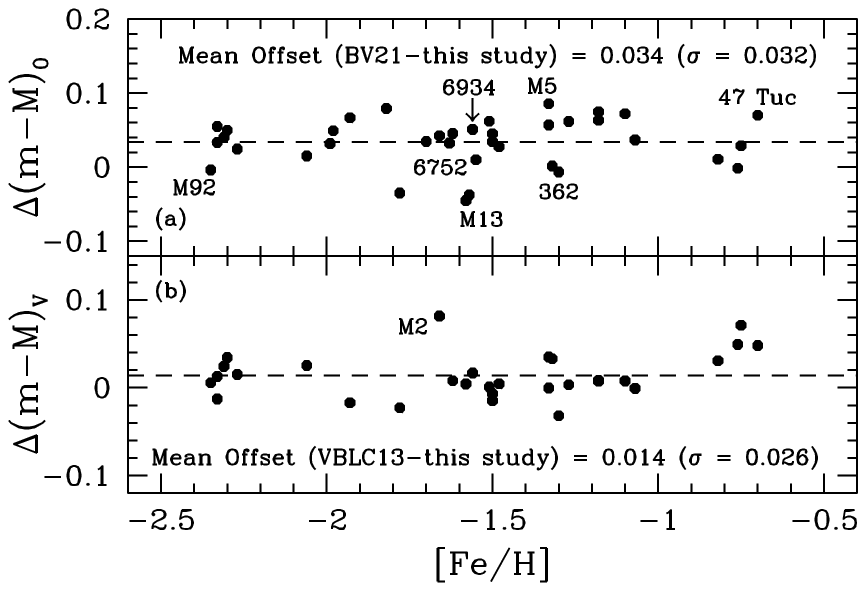}
\caption{Panel (a): Differences between the true distance moduli determined by
\citet{bv21} and the results of this study for the GCs that are listed in
Table~\ref{tab:t3}.  The few clusters that are explicitly identified are
discussed in the text.  Panel (b): Differences between the apparent distance
moduli derived by VBLC13 and the present findings.  The most discrepant results
were obtained for M$\,$2, which is explictly identified.}
\label{fig:f19}
\end{center}
\end{figure}

\citet[hereafter BV21]{bv21} have recently derived the distances to 162 GCs
by combining the results that they obtained using several methods with $\sim
1300$ published distance determinations.  Only a small fraction of their
findings are independent of the reddening; e.g., distances based on {\it Gaia}
EDR3 parallaxes.  In general, they adopted the $E(B-V)$ values from the papers
that were included in their survey or from the 2010 edition of the \citet{har96}
catalogue of cluster parameters.  However, they do not provide any reddening
information or their best estimates of $(m-M)_V$; consequently, there is no
other option but to compare $(m-M)_0$ values.  The uppermost panel of
Figure~\ref{fig:f19} shows that the true distance moduli that are implied by
the adopted reddenings and the derived $(m-M)_V$ values in Table~\ref{tab:t3}
for the 37 GCs that were considered in this investigation are smaller, in the
mean, by 0.034 mag than those reported by BV21 for the same clusters.  This
would imply an age difference of only $\approx 0.34$ Gyr if all other factors
that affect age determinations are unchanged.  As indicated in panel (a), the
standard deviation of the mean is 0.032 mag, which is comparable with the
estimated uncertainties of the ZAHB-based apparent distance moduli.

While this level of agreement is really quite good, the scatter in the results
is relatively large, especially at [Fe/H] $\sim -1.5$.  This could be a
reflection of the inhomogeneity of the BV21 findings.  For instance, both M$\,$5
and NGC$\,$362 have red HB populations that are easily matched to ZAHB models.
Whereas BV21 report a distance for NGC$\,$362 that is in excellent agreement
with the inferred value from appropriate ZAHB models, their distance for M$\,$5
implies that its HB stars are $\sim 0.08$ mag brighter than the same ZAHB.
This inconsistency is presumably the consequence of taking an average of many
distance determinations over the years that have employed a variety of methods
and made different assumptions concerning the cluster reddenings and chemical
abundances.  Indeed, prior to the revision of GC metallicities by CBG09, the
[Fe/H] values given by \citet{cg97}, which are higher by about 0.2 dex for many
clusters, were frequently adopted.  This would have the consequence of, e.g.,
increasing the distances that are derived via the classic MS-fitting technique.
The present work has the important advantage that the derived cluster properties
are very homogeneous as they are based on the same methods, the same ZAHB
models, and the same [Fe/H] scale (CBG09).   It is entirely possible that the
ZAHB-based distance moduli are too large or too small by a few hundredths of a 
magnitude, but the difference in the distance moduli of two clusters that have
almost the same metallicities and well-defined, nearly horizontal distributions
of HB stars, such as M$\,$5 and NGC$\,$362, should be a robust result.

\begin{figure*}
\begin{center}
\includegraphics[width=0.96\textwidth]{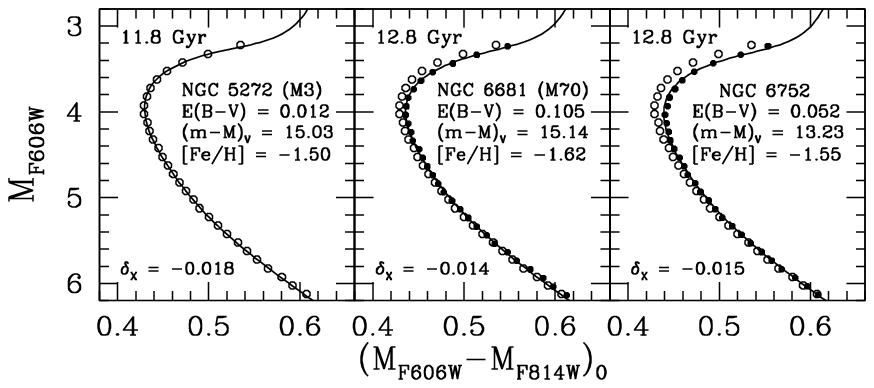}
\caption{Fits of isochrones for $Y = 0.26$ and the indicated ages and cluster
metallicities (from CBG09) to the median MS fiducial sequences of M$\,$3,
M$\,$70, and NGC$\,$6752 on the assumption of the specified cluster parameters.
The M$\,$3 CMD has been reproduced in the middle and right-hand panels to 
illustrate its location relative to those of the other two GCs.  The $\delta_X$
values specify the colour offsets that were applied to the isochrones in order
to obtain the fits to the upper-MS and turnoff observations that are shown;
these offsets are discussed in \S~\ref{sec:fits}.}
\label{fig:f20}
\end{center}
\end{figure*}

The $\Delta\,(m-M)_0$ values for NGC$\,$6934, NGC$\,$6752, and M$\,$13, which
have the same metallicities to within 0.03 dex according to CBG09, also span a
wide range in Fig.~\ref{fig:f19}a.  Of these three clusters, only the HB of
NGC$\,$6934 extends to very red colours, with the consequence that the fitting
of these stars to the relevant ZAHB, and the resultant value of $(m-M)_V$,
involve rather little uncertainty.  Fortunately, NGC$\,$6934 also has a
sufficient number of very blue HB stars that they may be easily compared with
similar stars in M$\,$13 on the $(M_{F336W}-M_{F606W})_0,\,M_{F606W}$ diagram.
As shown in Fig.~\ref{fig:f13}m, they superimpose one another very well if
M$\,$13 has $(m-M)_V = 14.45$ and $E(B-V) = 0.017$ (from dust maps).  It would
be possible to obtain the same distance modulus that was derived by BV21 
and a comparable match of the blue HB stars in M$\,$13 and NGC$\,$6934 if
M$\,$13 has $(m-M)_V = 14.40$ {\it and} $E(B-V) = 0.030$, but such parameter
values would present problems for the relative locations of the M$\,$13 and
M$\,$3 main sequences (see \S~\ref{sec:second}).  Note that the distance modulus
of NGC$\,$6752 as determined in this study and by BV21 differ by only 0.01 mag.

M$\,$92 and 47 Tuc have been explicitly identified in Fig.~\ref{fig:f19}a
because they stand apart from the other clusters in their respective metallicity
groups.  While an equally good fit of the HB stars in M$\,$92 to the ZAHB models
could be obtained on the assumption of a larger distance modulus by $\sim 0.03$
mag, which would move the filled circle that represents M$\,$92 close to the
points that represent the other GCs with [Fe/H] $\sim -2.3$, this would have
the consequence of making its MS appreciably redder at a given absolute 
magnitude than the MS fiducials of, e.g., M$\,$68, M$\,$30, and NGC$\,$5053.
This would be contrary to expectations if, as appears to be the case (see
\citealt{zmm23}), M$\,$92 has a somewhat higher median He abundance than the
other clusters.  Assuming an increased reddening to compensate for this
discrepancy would result in an unacceptable fit of the cluster HB stars to the
ZAHB models; recall the discussion of Fig.~\ref{fig:f6} in \S~\ref{subsec:lowz}.
Thus, provided that the ZAHB models are trustworthy, the apparent distance
modulus of M$\,$92 should be very close to $(m-M)_V = 14.72$, which is obtained,
in fact, if the distance of M$\,$92 is 8.501 kpc (from BV21) and it has a
reddening $E(B-V) = 0.022$ (from dust maps).

The previous section has already described the considerable evidence in 
support of $(m-M)_V = 13.30$ for 47 Tuc, but the main point that is relevant
for this paper is that this determination is favoured by ZAHB models when
current best estimates of the reddening and metallicity are assumed.  It is
also worth pointing out that, as reported by BV21, {\it Gaia} EDR3 parallaxes,
corrected for systematic errors, yield a distance of $4.367 \pm 0.18$ kpc,
which corresponds to $(m-M)_0 = 13.20 \pm 0.09$.  This is nearly identical
with the true modulus that is derived from $(m-M)_V = 13.30$ if the dust-map
reddening, $E(B-V) =0.030$, is adopted.  Unfortunately, there are only a few
GCs with sufficiently accurate parallax-based distances that yield true
distance moduli with uncertainties $\lta \pm 0.10$ mag.   In all such cases, the
ZAHB-based distance moduli derived in this study (and by BV21) agree with those
determinations to within their uncertainties.

The bottom panel of Fig.~\ref{fig:f19} shows that the apparent distance moduli
derived by VBLC13 differ from the present results by 0.014 mag, on average,
for 27 GCs in common to the two investigations.  As indicated, the standard
deviation of the mean is 0.026 mag.  Note that the plots provided by VBLC13
give the values of $(m-M)_{F606W}$ that were derived from fits of the cluster
HBs to ZAHB models; they were converted to $(m-M)_V$ using the $R_\lambda$
values given by \citealt{cv14}.  VBLC13 did not attempt to derive the apparent
distance moduli of several clusters with extremely blue HBs, such as NGC$\,$6397
and NGC$\,$6752, but instead determined their ages relative to those of clusters
with HBs that could be reliably fitted to ZAHB models using the so-called
``horizontal method" of determining relative cluster ages (see \citealt{vbs90}).
This makes use of the predictions from isochrones for the difference in colour
between the TO and the lower RGB as a function of age. 

It turns out that very close to the same ages are obtained when isochrones
are fitted to the turnoff observations on the assumption of ZAHB-based distance
moduli.  As shown in Figure~\ref{fig:f20}, M$\,$3 is predicted to have an age
near 11.8 Gyr, as compared with ages of $\approx 12.8$ Gyr for M$\,$70 and
NGC$\,$6752.  For the same three clusters, VBLC13 obtained ages of 11.75, 12.75,
and 12.50 Gyr, respectively, with $1\,\sigma$ uncertainties of $\pm 0.25$--0.38
Gyr.  Although not shown, similar plots were produced for 5 other clusters that,
like M$\,$70 and NGC$\,$6752, have HBs which are entirely to the blue of the
instability strip; except for M$\,$2, their ages also differ by $\le 0.3$ Gyr
from the determinations made by VBLC13.  Such differences are clearly too small to
significantly alter the GC age--metallicity relation that was derived by VBLC13
and discussed at some length by \citet{lvm13}.  This is to be expected given the
similarity of the derived distance moduli for most of the clusters in common to
the two studies (see Fig.~\ref{fig:f19}b).  It should be kept in mind,
however, that ages are quite sensitive to the assumed chemical abundances; e.g.,
DVKF17 obtained a higher age for M$\,$3 (12.6 Gyr) primarily because they
adopted lower values of [Fe/H] and [O/Fe].

The discordant results for M$\,$2 are mainly due to the difference in the
adopted values of $(m-M)_V$ values (see Fig.~\ref{fig:f19}b).  VBLC13 found a
larger distance modulus for M$\,$2 by about 0.08 mag because they adopted a
lower reddening ($E(B-V) = 0.046$ from the SDF98 dust maps).  Had this reddening
been assumed in the present study, the fitting of the cluster HB stars to the
ZAHB models would have yielded a distance modulus very similar to the value that
was derived by VBLC13.  However, it is not possible to obtain a satisfactory
superposition of the blue HB stars in M$\,$2 and M$\,$13 on the
$(M_{F336W}-M_{F606W})_0,\,M_{F606W}$ diagram if the cluster parameters given
by VBLC13 are adopted.  There is no such difficulty if M$\,$2 has $E(B-V) =
0.052$ and $(m-M)_V = 15.46$ (see Fig.~\ref{fig:f13}o).   This example
illustrates the valuable constraints on our understanding of GCs that are
provided by UV-optical CMDs.

VBLC13 also found somewhat larger distances moduli for the most-metal rich clusters
due to a preference for $(m-M)_V = 14.35$ for 47 Tuc from a consideration of its
eclipsing binary V69 and an assumption that mass loss along the RGB did not
exceed $0.20 \msol$.  Given the ambiguities of fitting the observed HBs to the
red ends of ZAHBs, VBLC13 acknowledge that their estimates of $(m-M)_{F606W}$
for clusters with [Fe/H] $\gta -0.8$ could easily be in error by $\pm 0.05$ mag.
The fact that the distance moduli derived in this paper for three of the four
metal-rich clusters are in good agreement with the distances reported by BV21
(see Fig.~\ref{fig:f19}a), which are based on a number of different methods,
indicates a preference for the present results for GCs with [Fe/H] $> -1.0$
over those given by VBLC13.

What is potentially one of the most important results of this investigation
is the discovery that the main sequences of many (most?) second-parameter
clusters, including M$\,$2, M$\,$12, M$\,$13, M$\,$70, NGC$\,$6752, and
NGC$\,$288, nearly coincide with, or are slightly redder than, the MSs of GCs
that have almost the same metallicities but much redder HB populations, such as
M$\,$3, M$\,$5, and NGC$\,$362, {\it if} dust-map reddenings and distance moduli
based on ZAHB models are adopted.  The same thing would be found for the
moderately reddened ($E(B-V) \approx 0.30$) clusters NGC$\,$5286 and
NGC$\,$5986, which also have extended blue HBs, if they have $E(B-V)$ values 
that are less than dust-map determinations by only $1.5\,\sigma$.  

Examples of these findings are presented in Fig.~\ref{fig:f20}, which shows that
the indicated cluster parameters lead to very slight redward offsets of the MSs
of both M$\,$70 and NGC$\,$6752, which have extended blue HBs, relative to the
M$\,$3 MS, even though the latter is the most metal-rich of the three GCs by
$\delta\,$[Fe/H] $\sim 0.1$ dex.  Provided that dust-map reddenings are
reliable, as they appear to be, all three clusters would appear to
have similar He abundances, in the mean, as otherwise the MSs of M$\,$70
and NGC$\,$6752 would be significantly bluer than that of M$\,$3.  Thus,
something other than helium abundance variations must be the primary cause of
the very different HB morphologies of clusters that comprise such famous
second-parameter pairs as M$\,$3--M$\,$13 and NGC$\,$288--NGC$\,$362 --- though
such variations appear to have produced the HBs of, e.g., M$\,$15 and
NGC$\,$2808.

Higher ages would tend to promote bluer HBs, though the difference in mass
associated with an age difference as high as 2--3 Gyr is much less than the
amount of mass that must be lost for stars with [Fe/H] $\sim -1.5$ to end up on
the blue HB after the He flash (recall Fig.~\ref{fig:f2}).  In any case, the
extensive surveys of GC ages by \citet{map09} and VBLC13 both concluded that
M$\,$3 and M$\,$13 are nearly coeval and that NGC$\,$288 is older than
NGC$\,$362 by $< 1$ Gyr.  Hence, age probably plays no more than a minor
role in accounting for the HB morphologies of second-parameter GCs.  The same
can be said of CNO since spectroscopic work has established that M$\,$3 and M$\,$13
have very similar C$+$N$+$O abundances (e.g., \citealt{ssb96}, \citealt{cm05}),
as do NGC$\,$288 and NGC$\,$362 (\citealt{cd88}). 

Because (i) mass loss is inconsequential for the CMD locations of MS stars, (ii)
stellar rotation can be expected to increase the mass-loss rates along the upper
RGB, and (iii) the frequency of rapid rotators has been observed to vary from
cluster to cluster (\citealt{prc95}, \citealt{bb03}), it would not be a surprise
if mass loss, facilitated by rotation, is mainly responsible for the observed
variations in the HB morphologies of clusters that have essentially the same
metallicities --- especially given the lack of viable alternative explanations.
This would mean that stellar rotation is the dominant second parameter.  Whether
or not the mean temperatures of the MS stars at a given absolute magnitude are
affected by rotation in some of the GCs that have very blue HBs is extremely
difficult to determine because such effects are likely to be smaller than those
arising from reddening and chemical composition uncertainties.  Nevertheless,
the present work does provide some tantalizing indications in support of this
possibility.

\section*{acknowledgements}
I am very grateful to Santi Cassisi for those ZAHBs which appear in
Fig.~\ref{fig:f1} that were generated using the BaSTI code, and for a number
of suggestions that have led to an improved paper.  Helpful comments and/or
useful information from Karsten Brogaard, Luca Casagrande, Pavel Denisenkov,
and Domenico Nardiello are also very much appreciated.

\section*{Data Availability}
The grids of ZAHB sequences and the means to interpolate in them to produce
individual ZAHB loci for specific values of [Fe/H], $Y$, and [O/Fe] within the
ranges for which they were computed (see \S~\ref{sec:models}) may be obtained
from https://www.canfar.net/storage/list/VRmodels.  The files relevant to this
investigation are: (i) vzahb.zip, which contains the ZAHB grids and various
Fortran codes to interpolate in these grids and to transform them to observed
CMDs, and (ii) README\_vzahb, which provides detailed instuctions on how to
use the computer programs that are provided.

\bsp   
\label{lastpage}

\end{document}